\documentclass[a4paper,11pt]{article}
\pdfoutput=1 
\usepackage{jinstpub} 
\usepackage{lineno}
\usepackage{multirow}
\usepackage{float}
\usepackage{comment}
\usepackage{bodegraph}
\usepackage[cp1252]{inputenc}
\usepackage{footnote}
\usepackage{mathrsfs}

\title{\boldmath Timing performances of front-end electronics with 3D-trench silicon sensors}

\author[a,1]{Gian Matteo Cossu}\note{Corresponding author.}
\author[a]{and Adriano Lai}

\affiliation[a]{Istituto Nazionale Fisica Nucleare, Sezione di Cagliari, Cagliari, Italy}

\emailAdd{gianmatteo.cossu@ca.infn.it}

\abstract{Detectors based on pixels with timing capabilities are gaining increasing importance in the last years. Next-to-come high-energy physics experiments at colliders require the use of time information in tracking, due to the expected levels of track densities in the foreseen experimental conditions. A promising solution to gain high-resolution performance at the sensor level is given by so-called 3D silicon sensors. The excellent intrinsic time resolution of a special case of 3D sensors, the trench type, is limited by residual non-uniformities in the duration of the induced currents. The intrinsic contribution of the sensor to the total time resolution of the system, when the detector is coupled to a front-end electronics, depends on the characteristics of the electronics itself and can be minimized with a proper design. This paper aims to analyze the possible performance in the timing of a typically-used front-end circuit, the Trans-Impedance Amplifier, considering different possible configurations. Evidence of the preferred modes of operation in sensor read-out for timing measurement will be given.
}


\arxivnumber{1234.56789} 

\begin{document}

\maketitle
\flushbottom

\newpage
\section{Introduction}
\label{sec:intro}

An important emerging requirement in experimental high-energy physics concerns the need of introducing time measurements at the level of the single pixel sensor. As an example, the Upgrade-II of the LHCb experiment at the CERN LHC, scheduled to take data in about a decade from now, has requirements of concurrent space and time resolutions of the order of 10~$\mu$m and at least 50~ps, respectively, per single pixel hit~\cite{LHCb-PII-Physics,FCC}. Such a trend is  foreseen to continue with even more severe requirements in the subsequent generation of collider experiments, where time resolutions in the range of 10-20 ps per hit will be necessary~\cite{FCC}. Radiation resistance against fluences approaching $10^{17}$~1-MeV~$n_{eq}/cm^2$ is also a fundamental requirement. 3D silicon sensors demonstrated the capability to satisfy all these extreme requirements at the same time~\cite{Brundu_2021,Andreaiworid}. A timing-optimized 3D sensor, designed with trench geometry (3D-trench in the following), was developed and produced within the TimeSPOT project \cite{timespot}. The induced-current signals produced in 3D-trench sensors, although extremely fast (typical charge collection time of 200 ps), have different duration depending on the position of the ionizing tracks with respect to the electrodes \cite{TCode,tcode2}. The effect on the output signals, once these current signals are processed by a specific electronics, can be analyzed with a simple model that takes into account the duration variations of the currents. This can be done by considering a specific circuit topology for the electronics. Usually, the signal produced by the sensor is processed by a trans-impedance amplifier, (TIA in the following), characterized by a time constant which is related to the bandwidth of the system. Once the type of electronics is chosen, it is possible to understand how the timing performance changes as a consequence of the different duration of the set of current signals that the system has to process.

In this respect, the present paper aims to describe the relationship between the so-called \emph{intrinsic} contribution of the sensor and the characteristics of the electronics, which affects the time constant of the system and also the electronic jitter. 

We start discussing some general concepts in section \ref{general}. Section \ref{sec:frontend} is dedicated to describe the typical circuit solution for reading a capacitive sensor (i.e the TIA amplifier), in two particular conditions defined according to the value of the time constant of the system with respect to the average duration of the currents in the sensor. Section \ref{sec:Time3d} is dedicated to analyse the intrinsic resolution of the 3D sensor in the two circuit conditions described in section \ref{sec:frontend} . Here we introduce the concept of \textit{timing propagation coefficient} $\mathscr{P}$ from sensor to electronics.
The propagation coefficient $\mathscr{P}$ allows to highlight the link between the intrinsic contribution of the 3D sensor and the characteristics of the front-end electronics. It is also effective to understand under which conditions the timing performance of the system can be improved.

\section{General contributions to the time resolution}\label{general}

When quoting the contributions to uncertainty in the measurement of time, the following main quantities are normally considered,
\begin{equation}\label{eq:sigmast}
   \sigma_{\rm t} = \sqrt {\sigma^2_{\rm tw} + \sigma^2_{\rm TDC} + \sigma^2_{\rm sens} +\sigma^2_{\rm ej}}~~,
\end{equation}

$\sigma_{\rm tw}$ (\emph{time-walk}) depends systematically on the fluctuations of the signal amplitude and can be corrected with dedicated signal processing techniques, as Constant Fraction Discrimination (CFD) or Leading Edge Discrimination (LED) with time-over-threshold measurement (TOT). $\sigma_{\rm TDC}$ depends on the digital resolution of the electronics (conversion error) and can be made negligible by proper design. The $\sigma_{\rm sens}$ term is what is commonly called the intrinsic resolution of the sensor and can depend on the effect of longitudinal non-uniformity in the energy deposit, due to delta rays, but also on differences in the signal shapes. The latter, in a 3D geometry, is mainly due to the different possible drift paths of the charge carriers in the sensor and produces variations in the signal current duration. This contribution depends only on the geometry of the sensitive volume and can be minimized if maximum uniformity in the electric field is obtained by sensor design~\cite{TCode,tcode2}. The $\sigma_{\rm ej}$ term (electronic jitter) depends on the front-end electronics rise time and signal-to-noise ratio (SNR) and together with $\sigma_{\rm sens}$ represents the main contribution to time resolution for 3D detectors. In order to obtain the best performance in terms of time resolution it is then necessary to minimize these two contributions:

\begin{equation}\label{eq:sigmast2}
   \sigma_{\rm t} \sim \sqrt {\sigma^2_{\rm sens} +\sigma^2_{\rm ej}}~~. 
\end{equation}

In the next sections we will deal with these two fundamental contributions of the time resolution, (i.e. $\sigma_{\rm sens}$ and $\sigma_{\rm ej}$), and how they change depending on the characteristics of the electronics.

\section{Front-end electronics for timing}
\label{sec:frontend}

The traditional textbook solution for the read-out of capacitive sensors is the well-known Charge Sensitive Amplifier (CSA), possibly followed by a suitable number of differentiating (CR) and integrating (RC) stages, realizing a so-called \emph{Shaper}~\cite{Spieler}. 

\begin{figure}[h!]
    \centering
    \includegraphics[width=0.45\textwidth]{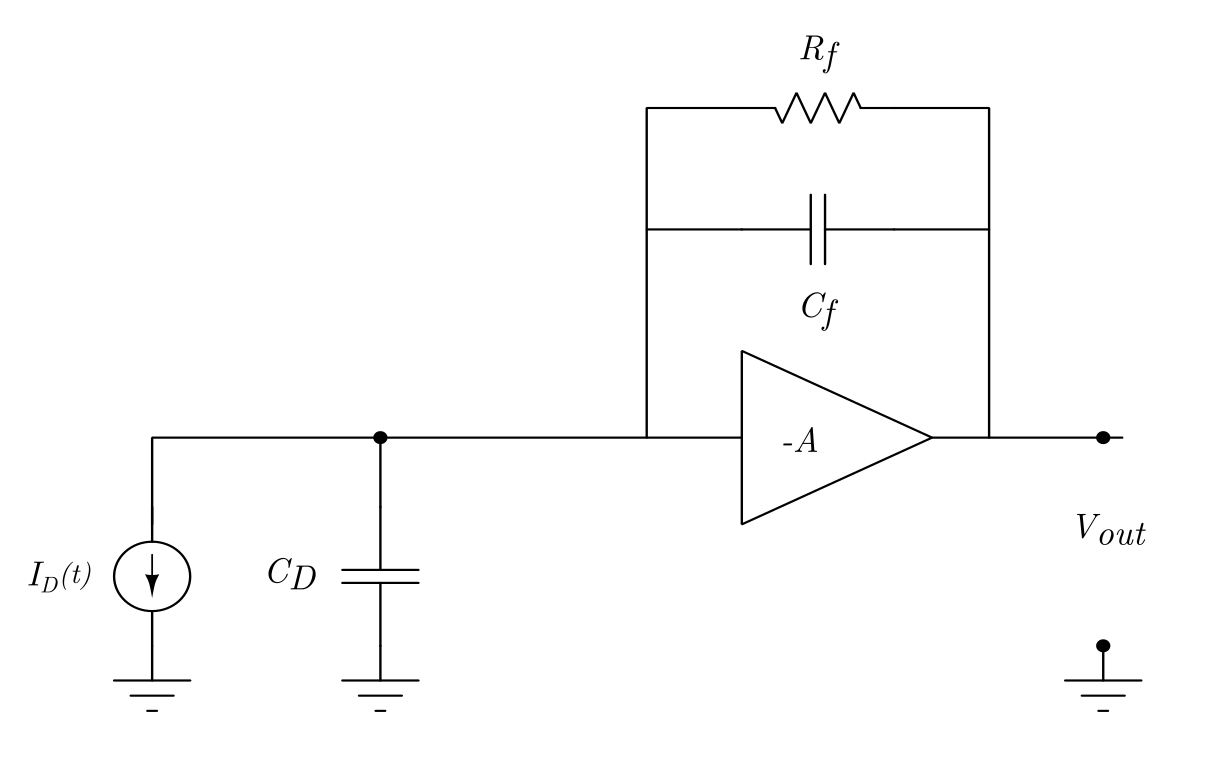}
    \includegraphics[width=0.45\textwidth]{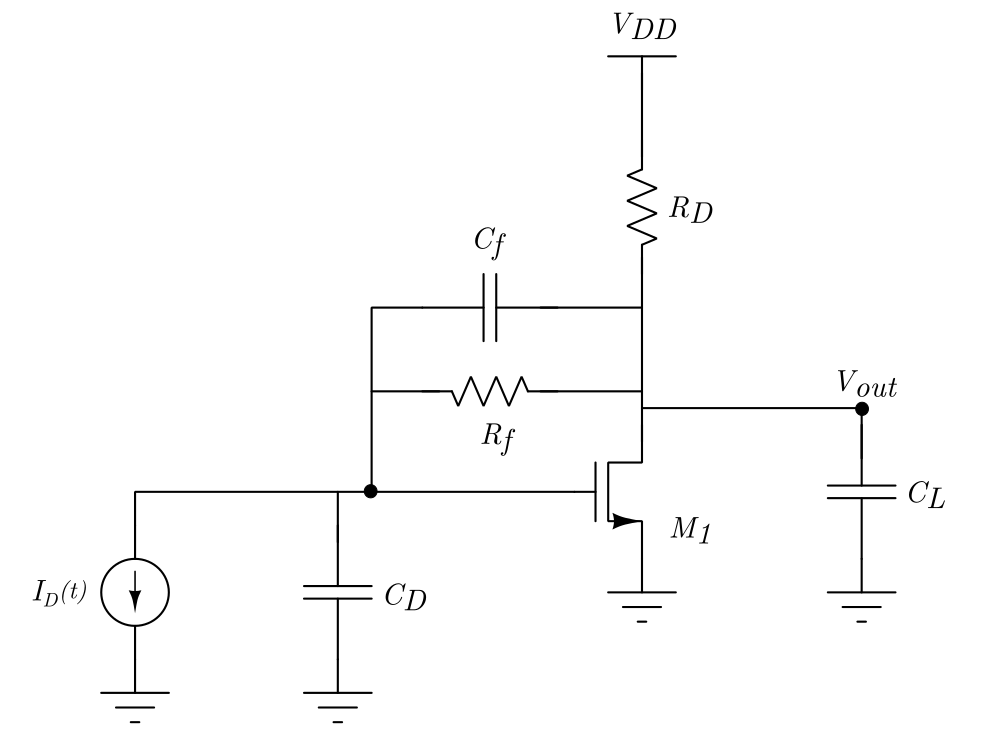}
    \caption{\footnotesize{General representation of the FB-TIA circuit (left). NMOS FB-TIA with a self-biased topology (right). The current generator $I_D(t)$ and the $C_D$ capacitance model the operation of the capacitive sensor.}}
    \label{fig:fbtia}
\end{figure}

Actually, the CSA circuit is a particular case of a more general configuration, that is the Trans-Impedance-Amplifier (TIA) with shunt-shunt feedback (FB-TIA), schematically shown in Fig.~\ref{fig:fbtia} (left). A simplified implementation of the TIA amplifier can be realized by a common-source NMOS in a so-called self-biased topology (Fig.~\ref{fig:fbtia}, right). 

The circuit (Fig.~\ref{fig:fbtia}, right) can be solved analytically  \cite{FastTiming} using the equivalent small parameters model, thus finding the following second order transfer function \footnote{This is the solution when the damping constant $\zeta=1$ as explained in \cite{FastTiming}.} \\
\begin{equation}
R_m(s)=\frac{R_fG_0}{1+G_0}\frac{(1-s\tau_z)}{(1+s\tau)^2}~,
\label{eq:FdT}
\end{equation}
where $ \tau_z=R^*C_f/G_0 $ is the time constant corresponding to the zero and $ G_0=(g_mR^*-\frac{R^*}{R_f}) $ is the DC gain ( $g_m$ is the trans-conductance of the transistor and $R^*= R_f || R_D$).

The time constant $\tau$, relative to the second-order pole reads

\begin{equation}
\tau \approx \sqrt{\frac{R_f\xi}{g_m}}~,
\label{omegan}
\end{equation}

and is dependent on the quantity $\xi$ that contains all the capacitances involved in the circuit,
\begin{equation}\label{xidef}
\xi = (C_LC_{in}+C_LC_f+C_{in}C_f)~.
\end{equation}

\begin{figure}[h]
\centering
	\includegraphics[width=0.8\textwidth]{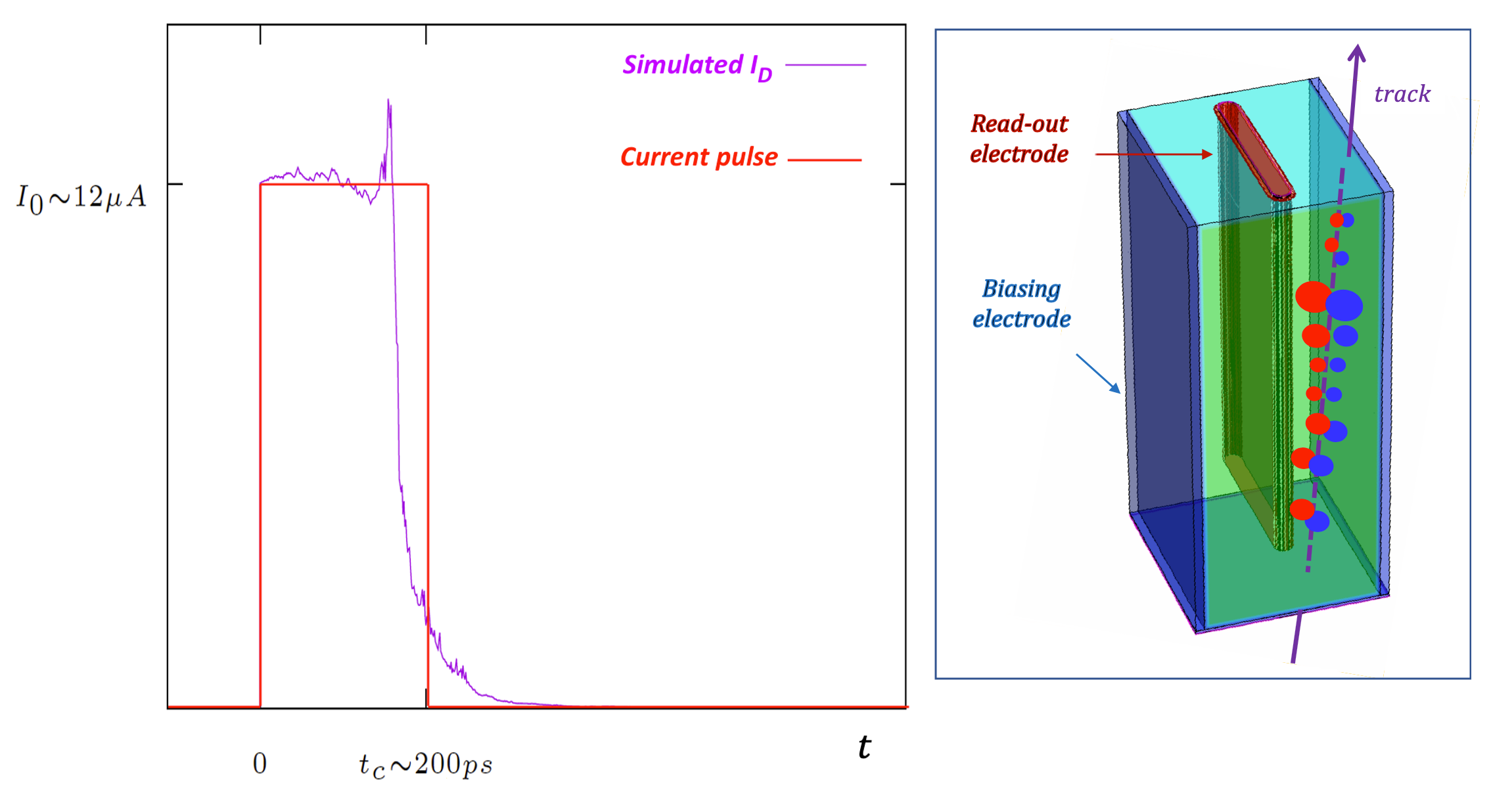}
    \caption{Current pulse $ I_D(t) $ (left) for a 3D pixel sensor with trench geometry (right). The simulated signal is obtained by TCoDe simulation~\cite{TCode,tcode2}. The sizes of the pixel are $ 55\times55\times150~\mu m^3 $.} 
    \label{fig:3Dresp}
\end{figure}

The trans-impedance in the \textit{s}-domain $R_m(s)$ (Eq.~\ref{eq:FdT}) needs to be convoluted with the sensor current signal $ I_D(s) $, in order to get the output voltage of the circuit. We consider here the simplified condition of a 3D-trench sensor (Fig.~\ref{fig:3Dresp}), operating with charge carriers both at saturation velocities. In this case, the current has a shape that can be modelled as a simple rectangular pulse, having a width of duration $ t_c $ and an amplitude $ I_0 $, such that the product $ I_0 \cdot t_c $ equals the total charge $ Q_{in} $ deposited by the particle (Fig.~\ref{fig:3Dresp}). The time $ t_c $ is the \emph{charge collection time} and is defined as the time required for all charge carriers to reach their respective electrodes and stop inducing.  This rectangular-shaped description does not take into account the different drift velocities of the carriers, but it is still a more realistic description compared to describing the current pulse as a pure Dirac delta function. The current  can then be expressed in the \textit{s}-domain as

\begin{equation}
I_D(s)=I_0\frac{1-e^{-st_c}}{s}~,
\end{equation}

so that the output voltage $ V_{out}(s) $ can be written as

\begin{equation}\label{csa_vout}
V_{out}(s)=I_0\frac{1-e^{-st_c}}{s}\frac{R_fG_0}{1+G_0}\frac{(1-s\tau_z)}{(1+s\tau)^2}~.
\end{equation}

Taking the inverse Laplace transform in the time domain we have the signal\\
\begin{equation}
V_{out}(t)=\mathcal{L}^{-1}(t)  \Bigg\{I_0\frac{1-e^{-st_c}}{s}\frac{R_fG_0}{1+G_0}\frac{(1-s\tau_z)}{(1+s\tau)^2} \Bigg\}~,
\end{equation}

which corresponds to a voltage signal described by the function
\begin{multline} \label{vout_sol}
V_{out}(t)=I_0\frac{R_fG_0}{1+G_0} \Bigg\{ \Bigg[1-e^{-\frac{t}{\tau}}\Bigg(1+\frac{t}{\tau}\Bigg(1+\frac{\tau_z}{\tau}\Bigg)\Bigg)\Bigg] - \\ \theta(t-t_c)\Bigg[1-e^{-\frac{(t-t_c)}{\tau}}\Bigg(1+\frac{(t-t_c)}{\tau}\Bigg(1+\frac{\tau_z}{\tau}\Bigg)\Bigg)\Bigg] \Bigg\}~.
\end{multline}

It is interesting to analyze the behaviour of the TIA circuit in two different operating conditions, distinguished by the size of the circuit time constant $\tau$ with respect to the duration of the input current: the \textit{charge collection time} $t_c$. This is accomplished in the next two subsections.

\subsection{Condition I: Charge Sensitive TIA \texorpdfstring{($\tau \gg  t_c$)}{}} 

This condition (named CS-TIA in the following) is typical of a CSA-based input stage, where the value of the feedback resistor $ R_f $ is maximized to have a better SNR. This is an optimal configuration when the precision in the signal amplitude measurement is more important than preserving the signal speed and time resolution. In any case, the use of the CSA configuration often remains a convenient compromise between overall performance and power consumption. The bandwidth of the TIA is kept much smaller compared to the bandwidth of the current pulse and, consequently, the shape of the current signal is not preserved. With a given trans-conductance $ g_m $ of the input transistor, the output voltage reaches quickly the maximum achievable slope, to decrease then exponentially with time. Since we have the factor $\theta(t-t_c)$ in the solution \ref{vout_sol}, we can consider two cases:\\

when $ t<t_c $ we get the output signal

\begin{equation}
V_{out}(t)_{t<t_c}=I_0\frac{R_fG_0}{1+G_0} \Bigg\{ \Bigg[1-e^{-\frac{t}{\tau}}\Bigg(1+\frac{t}{\tau}\Bigg(1+\frac{\tau_z}{\tau}\Bigg)\Bigg)\Bigg]~,
\label{voutCSA}
\end{equation}\\

when $ t>t_c $ the output signal expression becomes

\begin{equation}
V_{out}(t)_{t>t_c}= I_0\frac{R_fG_0}{1+G_0}  e^{-\frac{t}{\tau}}\Bigg(\frac{t}{\tau}(e^{\frac{t_c}{\tau}}-1)(\frac{\tau_z+\tau}{\tau})+\frac{e^{\frac{t_c}{\tau}}(\tau^2-t_c(\tau-\tau_z))}{\tau^2}-1\Bigg)~.
\label{eq:VoutPos}
\end{equation}

We can then define

\begin{equation*}
A=I_0\frac{R_fG_0}{1+G_0}~~~;~~~B=(e^{\frac{t_c}{\tau}}-1)  \Bigg(\frac{\tau_z+\tau}{\tau}\Bigg)~~~~;~~~C=\frac{e^{\frac{t_c}{\tau}}(\tau^2-t_c(\tau-\tau_z))}{\tau^2}-1~~.
\label{ABCdef}
\end{equation*}

Therefore, equation~\ref{eq:VoutPos} can be written as
\begin{equation}
V_{out}(t)_{t>t_c}=Ae^{-\frac{t}{\tau}} \Bigg(B\frac{t}{\tau} + C\Bigg)
\label{eq:VoutCSA1}~.
\end{equation}

\begin{figure}[H]
\includegraphics[width=0.95\textwidth]{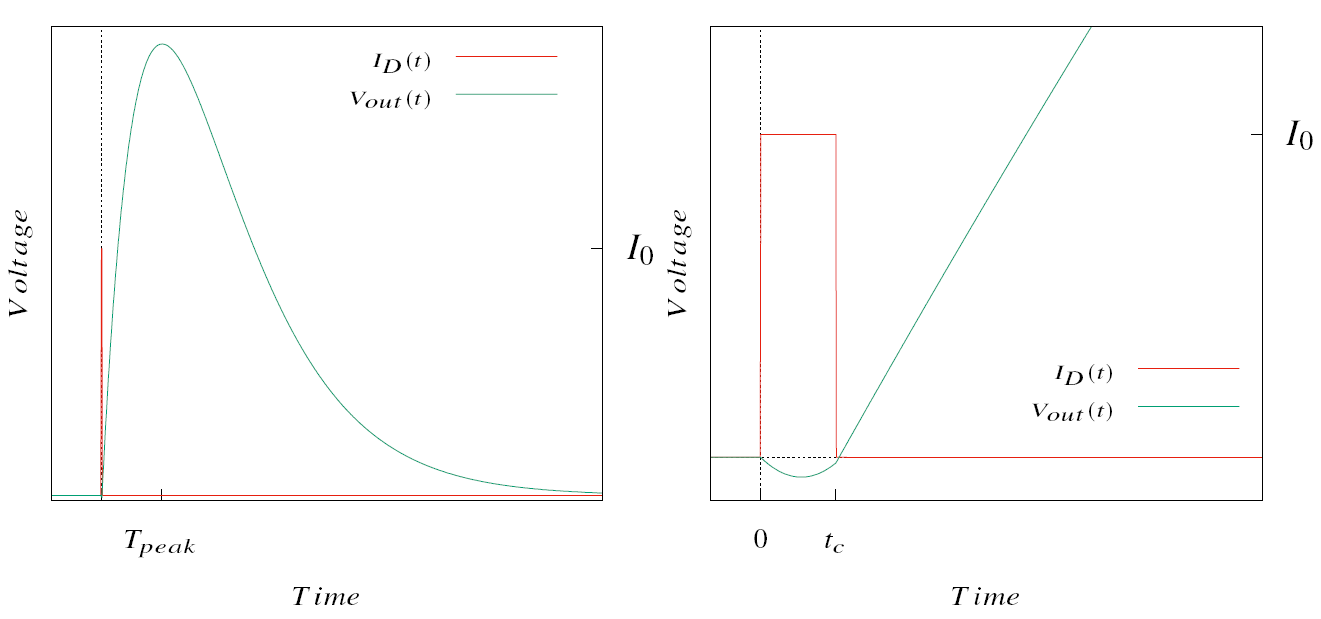}
	\caption{\footnotesize {Calculated output voltage $ V_{out} $ in the CS-TIA configuration. Left: Due to the high peaking time ($T_{peak} \approx \tau$) with respect to average collection time, the current signal can be approximated by a Delta function. Right: Detail of the under-shoot during $t < t_c$}.}
	\label{fig:VoutCSA}
\end{figure}

The peaking time $ T_{peak} $ and voltage peak $ V_{peak} $ are
\begin{equation}
T_{peak}=\frac{B-C}{B}\tau~ = 
\frac{e^{\frac{t_c}{\tau}}(\tau_z(\tau-t_c)+\tau t_c)-\tau \tau_z}{(\tau_z+\tau)(e^{\frac{t_c}{\tau}}-1)}~;
\label{tpeakComplete}
\end{equation}
\begin{equation}
V_{peak}=I_0\frac{R_fG_0}{1+G_0}e^{-\frac{T_{peak}}{\tau}}B \approx  \frac{Q_{in} \cdot R_f}{e \tau}~.
\label{VpeakCS}
\end{equation}
The output signal $V_{out}$ is shown in Fig.~\ref{fig:VoutCSA}. During charge collection, the signal starts having negative values. For $t>t_c$ the signal is positive with a positive derivative, reaching a maximum at $T_{peak}\approx \tau$. We can anticipate here that this condition does not appear as the best possible one when the speed of the sensor is to be fully exploited.

\begin{figure}[H]
\centering
\includegraphics[width=0.9\textwidth]{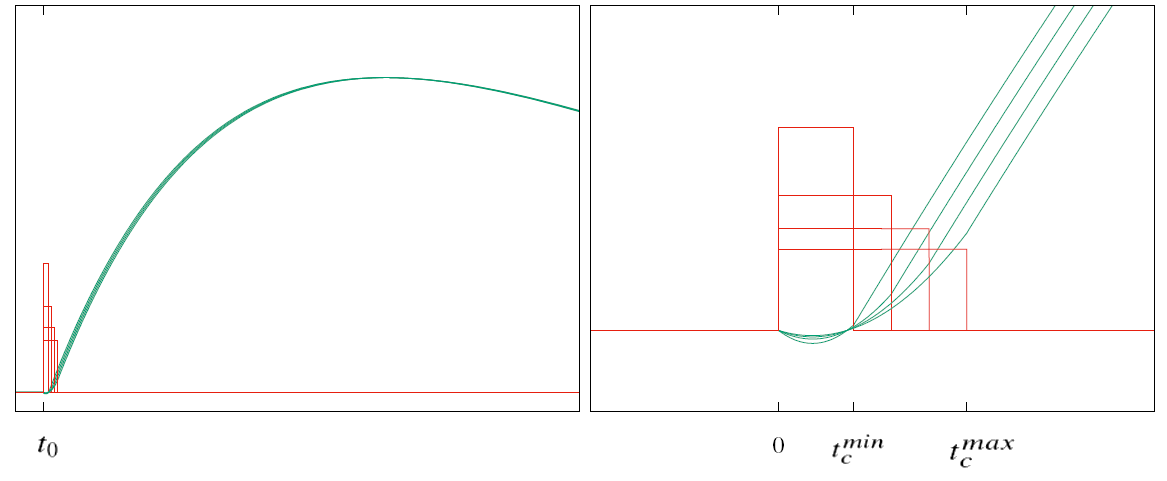}
	\caption{\footnotesize {Output voltage $ V_{out} $ in the CS-TIA configuration for current pulses with different duration $t_c$ and same charge $Q_{in}=I_0\cdot t_c$ (left). Detail of the under-shoot and slope of the signal for different values of $t_c$ (right)}}
	\label{fig:VoutCSA2}
\end{figure}

If the duration of the induced current signal is much shorter compared to the time constant ($\tau\gg t_c$), the slope of the voltage output signal becomes almost independent of $t_c$. As shown in Fig.~\ref{fig:VoutCSA2}, different charge collection times lead to a delay of the output signal, but the initial slope stays approximately constant. The maximum slope for every current is reached at time $t_c$ and then decreases exponentially. The maximum signal slope is

\begin{equation}
\frac{dV}{dt}\approx \frac{Q_{in} \cdot g_m}{\xi},
\label{eq:dVdtCSA}
\end{equation}

where $Q_{in}$ is the total charge (for the rectangular pulse $Q_{in}=I_0 t_c$), while $\xi$ has been defined in Eq. \ref{xidef}.

\subsection{Condition II: Fast-TIA \texorpdfstring{($\tau \approx t_c $)}{}}

\begin{figure}[h]
	\centering
	\includegraphics[scale=0.39]{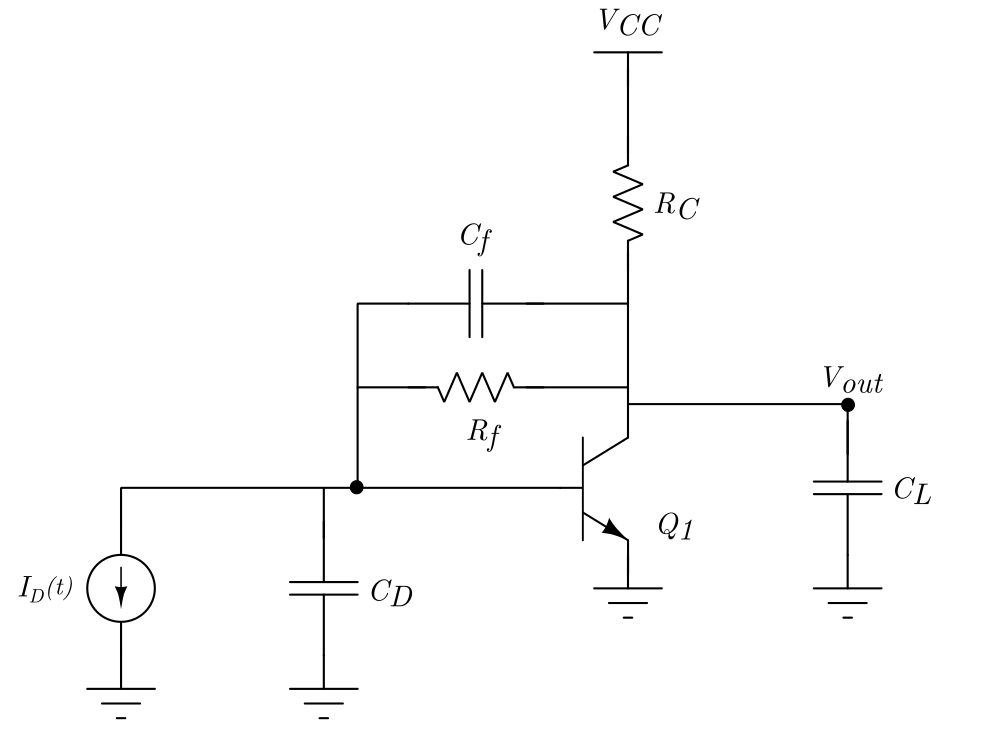}
	\includegraphics[scale=0.39]{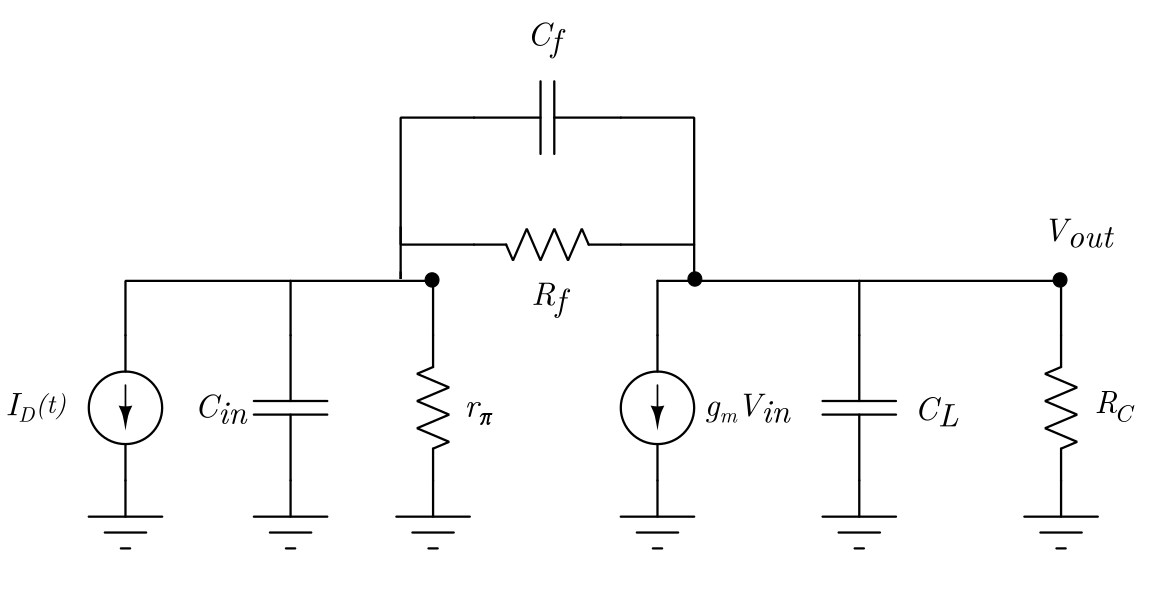}
	\caption{\footnotesize{Schematic of the FB-TIA with bipolar transistor NPN} (left) and corresponding small signal model (right).}
	\label{fig:SiGe}
\end{figure}

A possible solution for realizing a FB-TIA having a time constant of the same order of the charge collection time $ t_c $ is using the same self-biased scheme of Fig.~\ref{fig:fbtia} implemented with a high-bandwidth Si-Ge bipolar transistor stage, as illustrated in Fig.~\ref{fig:SiGe}. It is thus possible to take advantage of the benefits of the Si-Ge devices that allow reaching small input capacitances and high-frequency transitions of the order of 100 GHz also for discrete-component circuit solutions.

Also in this case the circuit can be solved analytically \cite{FastTiming}, finding the following transfer function
\begin{equation}
R_m(s) = \frac{R_{m_0}}{(1+s\tau)^2}~,
\label{eq:RmFast}
\end{equation} 

where $R_{m_0}$ is the DC-Transimpedance
\begin{equation}\label{key}
R_{m_0}=\frac{r_\pi g_mR_C R_f-r_\pi R_C}{(R_f+R_C+r_\pi(1+g_mR_C)},
\end{equation}

and the time constant is
\begin{equation}\label{tau_FT}
\tau=\sqrt{\frac{R_fR_C r_\pi \xi}{r_\pi(1+g_mR_C)+R_C+R_f}} \approx \sqrt{\frac{R_{m_0}\xi}{g_m}}~.
\end{equation}

The output voltage is given by the convolution with the current pulse and reads
\begin{equation}
\resizebox{.85\hsize}{!}{$
	V_{out}(t)=I_0R_{m_0} \Bigg\{ \Bigg[1-e^{-\frac{t}{\tau}}\Bigg(1+\frac{t}{\tau}\Bigg)\Bigg] - \theta(t-t_c)\Bigg[1-e^{-\frac{(t-t_c)}{\tau}}\Bigg(1+\frac{(t-t_c)}{\tau}\Bigg)\Bigg] \Bigg\}$}~.
	\label{eq:VoutBJT}
\end{equation}

Considering separately the two signals, with respect to the charge collection time $t_c$,
\begin{equation}\label{eq:voutFast}
V_{out}(t) = \begin{cases} 
I_0R_{m_0}(1-e^{-\frac{t}{\tau}}(1+\frac{t}{\tau})) & \mbox{if } t<t_c, \\
\\
I_0R_{m_0}e^{-\frac{t}{\tau}}( B\frac{t}{\tau}+C) & \mbox{if } t>t_c, \end{cases}
\end{equation}

where
\begin{equation*}
B=e^{\frac{t_c}{\tau}}-1  ~~~;~~~C=e^{\frac{t_c}{\tau}}\Big(1-\frac{t_c}{\tau}\Big)-1~~.
\label{ABCdef}
\end{equation*}

Taking the derivative of the voltage signal for $t<t_c$ we can find the slope of the signal
\begin{equation}\label{Vprimet}
	V_{out}^{'}(t)_{t<t_c}=\frac{I_0R_{m_0}}{\tau^2}e^{-\frac{t}{\tau}}t~.
\end{equation}

The slope is maximum at $t=t_c$, so we can write

\begin{equation}\label{eq:dVoutfast}
	V_{out}^{'}(t_c)=\frac{Q_{in}R_{m_0}}{\tau^2 e}~.
\end{equation}

The peaking time $ T_{peak} $ is 
\begin{equation}\label{Tpeak_fast}
T_{peak}=\frac{e^{\frac{t_c}{\tau}}t_c}{e^{\frac{t_c}{\tau}}-1}~.
\end{equation}

\begin{figure}[h]
	\includegraphics[scale=0.39]{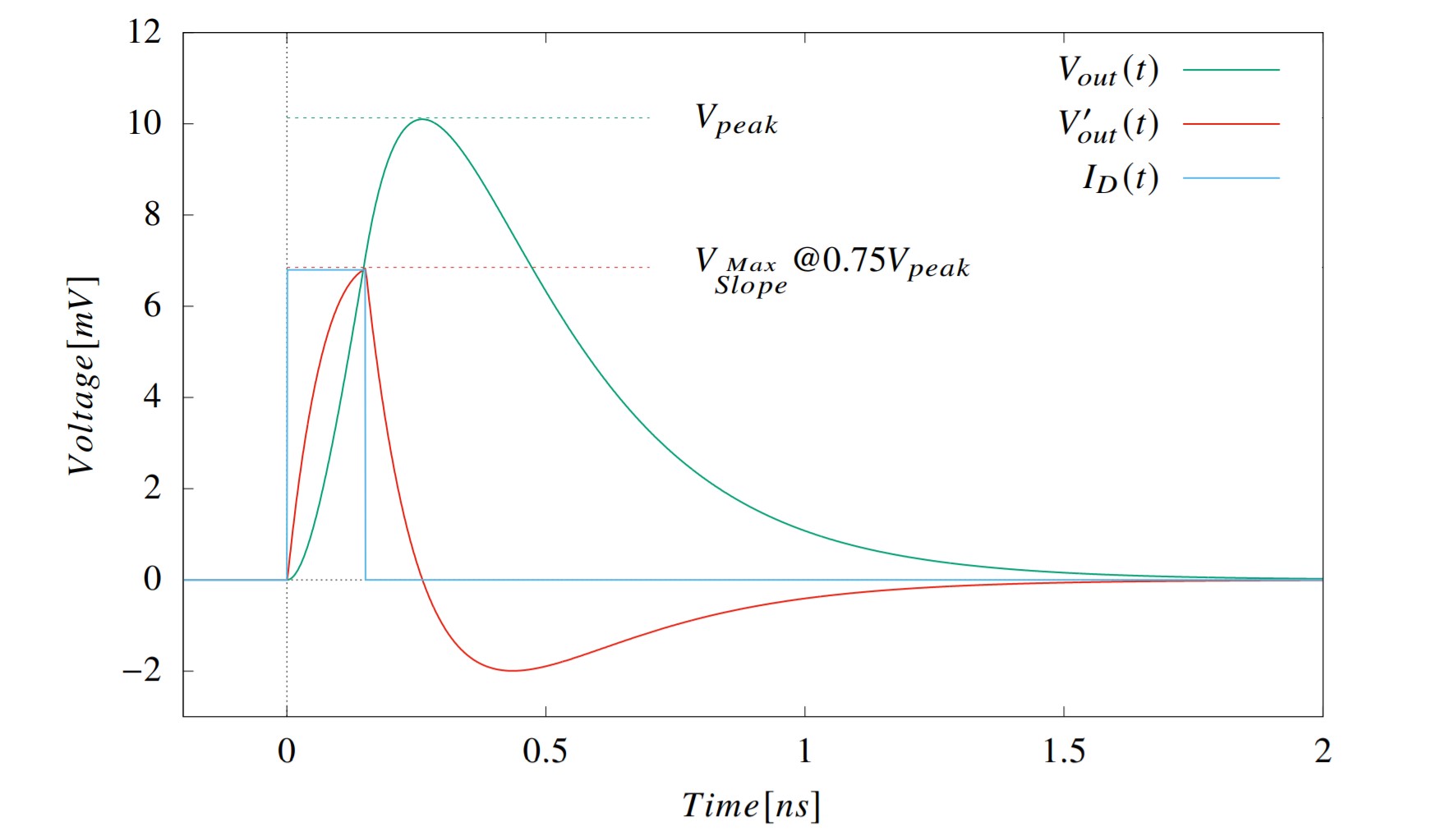}
\caption{\footnotesize{Example of the output signal and its derivative for a Fast-TIA.}}
\label{fig:fastTIAsig}
\end{figure}

The derivative is shown in Fig.~\ref{fig:fastTIAsig}. Also in this case the maximum value for the slope is reached at time $t_c$ but here we are at a much higher fraction of the voltage signal. In particular,

\begin{equation}\label{vpeakft}
V_{peak} = V_{out}(T_{peak})_{t>t_c} = I_0 R_{m_0} e^{-\frac{e}{e-1}}(e-1) \approx 0.353 \cdot I_0 R_{m_0}~.
\end{equation}

The voltage at time $ t=t_c$, that is the maximum slope voltage, is
\begin{equation}\label{eq:VoutfastmaxSl}
V_{\substack {\scriptscriptstyle Max\\Slope}}=I_0 R_{m_0}{\big(\frac{e-2}{e}\big)}\approx 0.264 \cdot I_0 R_{m_0}~. 
\end{equation}

Taking the ratio between $ V_{\substack {\scriptscriptstyle Max\\Slope}} $ and $ V_{peak} $ we find
\begin{equation}\label{key}
V_{\substack {\scriptscriptstyle Max\\Slope}} \approx 0.748 \cdot V_{peak}~.
\end{equation}

In the condition of $ \tau \approx t_c $, the maximum slope is reached at 
about $ 75\% $ of the peaking value (Fig.~\ref{fig:fastTIAsig}).

\section{Intrinsic time resolution of 3D-trench sensor with different front-end electronics}
\label{sec:Time3d}

In the present section, we analyze the performance in time resolution of the two configurations analysed above, that is the CS-TIA ($\tau \gg  t_c$) and the Fast-TIA ($\tau \approx t_c$) configurations. It is of particular interest to study how the characteristics of the electronics affect the \emph{primitive} sensor time resolution. The latter is usually referred to as \textit{intrinsic resolution}, and is strictly related to the spread of the charge collection time distributions (CCT), which in 3D sensors can be relatively small (standard deviations in the range of tens of ps).

\subsection{Sensor contribution to the time resolution}
\label{subsec:SensorJitter}

\begin{figure}[h]
	\centering
	\includegraphics[scale=0.5]{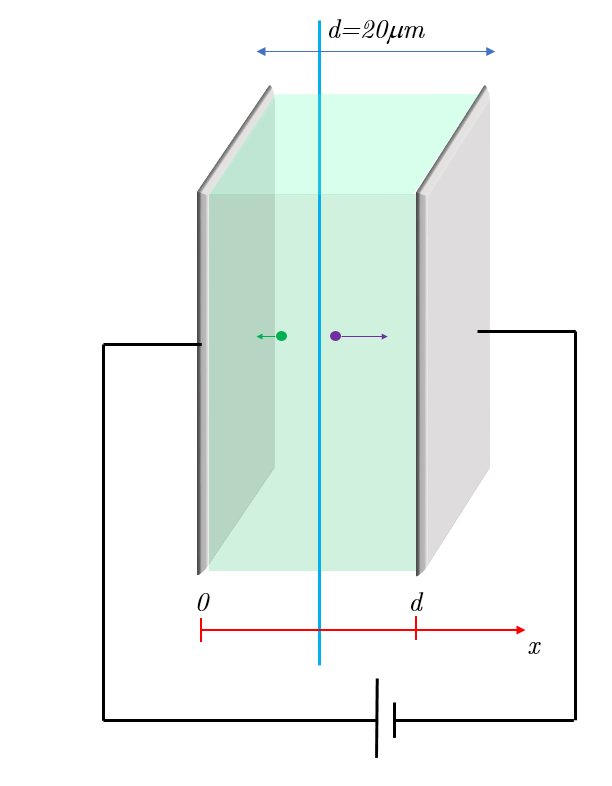}
	\includegraphics[scale=0.38]{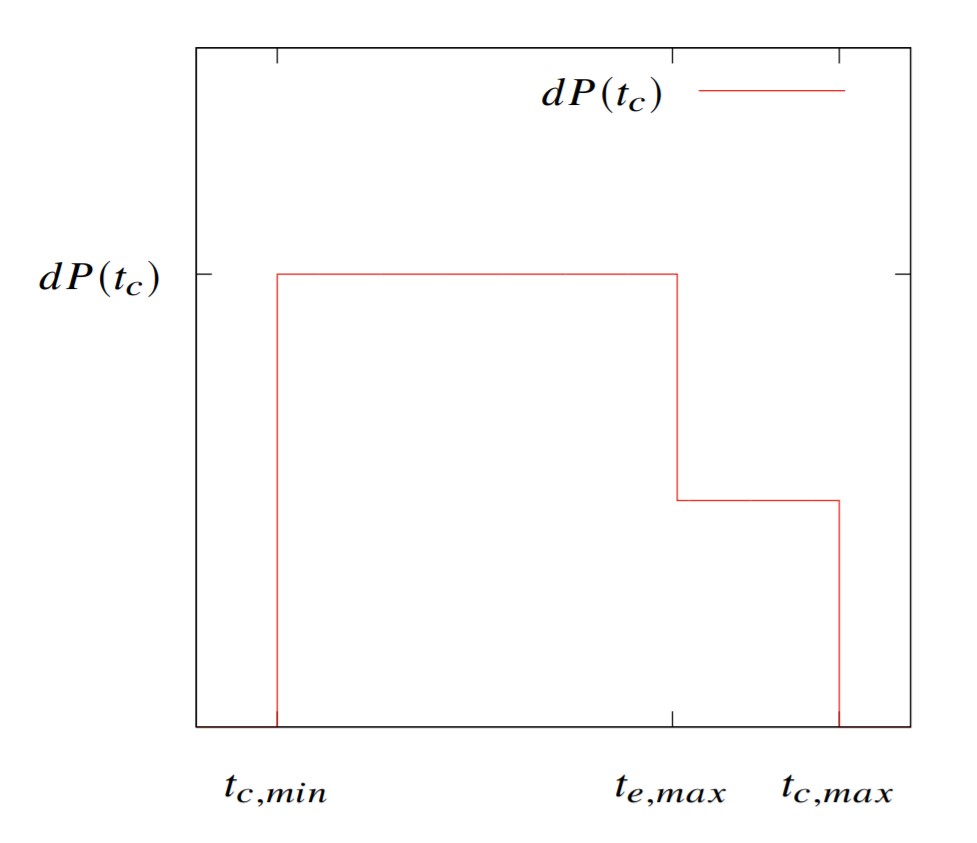}
	\caption{\footnotesize{Ideal parallel-plate sensor with 3D geometry (left), Probability Distribution Function of Charge collection time distribution (right)}}
	\label{fig:ideal3D}
\end{figure}

We consider an ideal 3D geometry with flat parallel-plate electrodes (Fig.~\ref{fig:ideal3D}). This choice is motivated by the high intrinsic speed of this kind of geometry~\cite{JINST-TimeSpot}. In this case $ t_c $ depends on the hit position of the impinging particle. For tracks closer to the electrode at the higher potential, the sensor will collect electrons very quickly while holes would induce for a longer time since they have to travel a longer distance and they move slower. The same argument can be used in the opposite case with tracks close to the electrode at lower potential with a short current signal from holes and a longer pulse for electrons. 
\begin{figure}[h]
	\centering
	\includegraphics[scale=0.35]{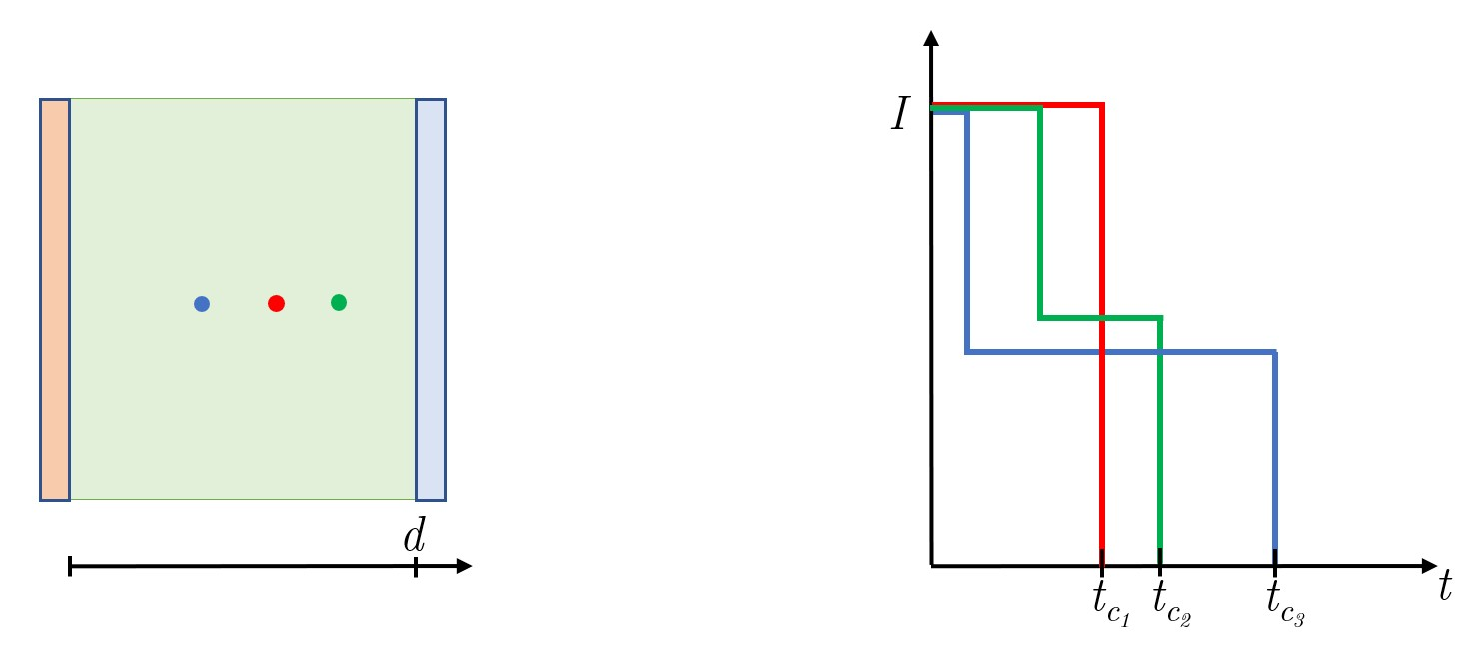}
	\caption{\footnotesize{Sensor with 3D geometry from the top, the colored dots represent the point where the track has passed (left), corresponding ideal current to the deposit on the sensor (right)}}
	\label{fig:ideal3D1}
\end{figure}

The shape of such current signals (Fig.\ref{fig:ideal3D1}, right) is very similar to simple rectangular pulses with the difference that, since the two carriers typically induce for different times, when one has finished inducing, we get a drop in the current amplitude. In the case where the charge collection time is minimum we get exactly a rectangular pulse.

We can assume for now that this difference in shape does not affect very much the obtainable intrinsic resolution, that will be instead dominated by the dispersion in the current signal duration $t_c$. This will allow us in the following to calculate the obtainable intrinsic resolution using the voltage signal expression derived in the previous sections. 

When we have electric fields strong enough for both charge carriers to reach the respective saturation velocities $ v_e $ and $ v_h $, the electrons and holes speeds in silicon become similar and are the same throughout the sensor. This is equivalent to neglecting weighting field non-uniformity. We will have a minimum $ t_c^{min} $ and a maximum $ t_c^{max} $ for the charge collection time $ t_c $ (see Fig.\ref{fig:VoutCSA2} right). Assuming that the distance between the electrodes is $ d=20~\mu m $ and using reasonable values for silicon saturation velocities, we obtain

\begin{equation}
t_c = \begin{cases}  t_c^{min}=\frac{d}{v_e+v_h}\sim 100~\text{ps} & \mbox{if } x=\frac{v_e d}{v_e+v_h} \\
\\
t_c^{max}=\frac{d}{v_h}\sim 210~\text{ps} & \mbox{if } x=d \end{cases}~~
\end{equation}

The current signal duration $t_c$ of the detector populate a distribution that has an average $\overline{t_c}$ and a certain standard deviation $\sigma_{t_c}$. 
For simplicity we can assume that the charge collection times generated spanning the detector width are all equally probable, (actually, shorter charge collection times are more probable, as electrons move faster,  Fig.~\ref{fig:ideal3D} right). Following such an assumption, both charge carriers have the same saturation velocity and we obtain a rectangular distribution corresponding in the ideal case to a dispersion

\begin{equation}\label{key}
\sigma_{t_c}\sim \frac{t_c^{max}-t_c^{min}}{\sqrt{12}}\approx 32~\text{ps}.
\end{equation}

This is not yet the intrinsic resolution of the sensor, but we will see in the next subsections that the final resolution for 3D-trench sensors is strongly related to the standard deviation $\sigma_{t_c}$ through a well defined \emph{propagation coefficient}.


\subsection{Intrinsic time resolution for different discrimination algorithms}
\label{subsec:FEJitter}

We analyze here the effect of the discrimination stage in the two cases of CS-TIA and Fast-TIA. We consider two kinds of typical discrimination techniques: LED and CFD. We limit our discussion to the intrinsic contribution due to the sensor, considering the \emph{time-walk} for the LED case, as a mere systematic (and processing-recoverable) effect.


\subsubsection{The timing Propagation Coefficient \texorpdfstring{$\mathscr{P}$}{}}
\label{subsubsec:leCSTIA}

Let us consider the case of a fixed voltage threshold set at voltage $ V_{th} $. The threshold needs to be at a higher voltage of the noise floor to avoid spurious hits, which in the case of the CS-TIA means considering the solution for the voltage signal at $t>t_c$. In principle, the threshold time $t_{s}$ is the time that satisfies the equation

\begin{equation}\label{le_1}
V_{th}=V_{out}(t_{s})~.
\end{equation}

We want to understand how this time $t_{s}$ is affected by the fact that the sensor produces a population of current signals with different charge collection times $t_c$. This is equivalent to consider $t_{s}$ as a function of the charge collection time $t_c$. Assuming Eq. \ref{le_1} could be solved to find the time $t_{s}$, the fluctuation of the time at the threshold can then be obtained by the error propagation formula

\begin{equation}
\sigma_{t_{s}}=\bigg(\frac{\partial t_{s}}{\partial t_c}\bigg)\sigma_{t_c}~.
\end{equation}

The proportionality factor between the resolution at the threshold $\sigma_{t_{s}}$ and the standard deviation of the CCT distribution $\sigma_{t_{c}}$, can be defined as the \textit{timing propagation coefficient} $\mathscr{P}$ from sensor to electronics:

\begin{equation}
\mathscr{P}=\bigg(\frac{\partial t_{s}}{\partial t_c}\bigg)~.
\end{equation}

The introduction the propagation coefficient $\mathscr{P}$ can help us to understand how the CCT distribution is decisive about the final performance of the system. In the following, we are interested to see how $\mathscr{P}$ changes for the two cases: CS-TIA and Fast-TIA.

\subsubsection{Constant-fraction time resolution in CS-TIA}

When using a CFD, the threshold is always set at the same fraction of the maximum value of the output voltage. This allows us to correct the time-walk given by the signal with the same charge collection times but different charges. In the CS-TIA case, being the electronics slow with respect to the duration of the pulse, this correction is essential to have good performance, otherwise the time resolution would strongly worsen. If a leading edge discrimination is used instead, this needs to be supported by a dedicated processing to compensate for the time-walk fluctuations, such as measuring the time over threshold (TOT), which can allow to reach an equivalent performance as the CFD. For this reason, we limit the calculation of the propagation coefficient in the CS-TIA case only to the CFD case, showing that the obtainable intrinsic resolution is independent of the chosen threshold. The voltage at the threshold can be written as
\begin{equation}\label{CFD_cs}
V_{out}(t_s)_{t>t_c}=\alpha V_{out}(T_{peak})~,
\end{equation}

$ t_s $ is the time at threshold, set at the fraction $ \alpha $ of the voltage peak.\\

\begin{figure}[h]
	\centering
	\includegraphics[scale=0.30]{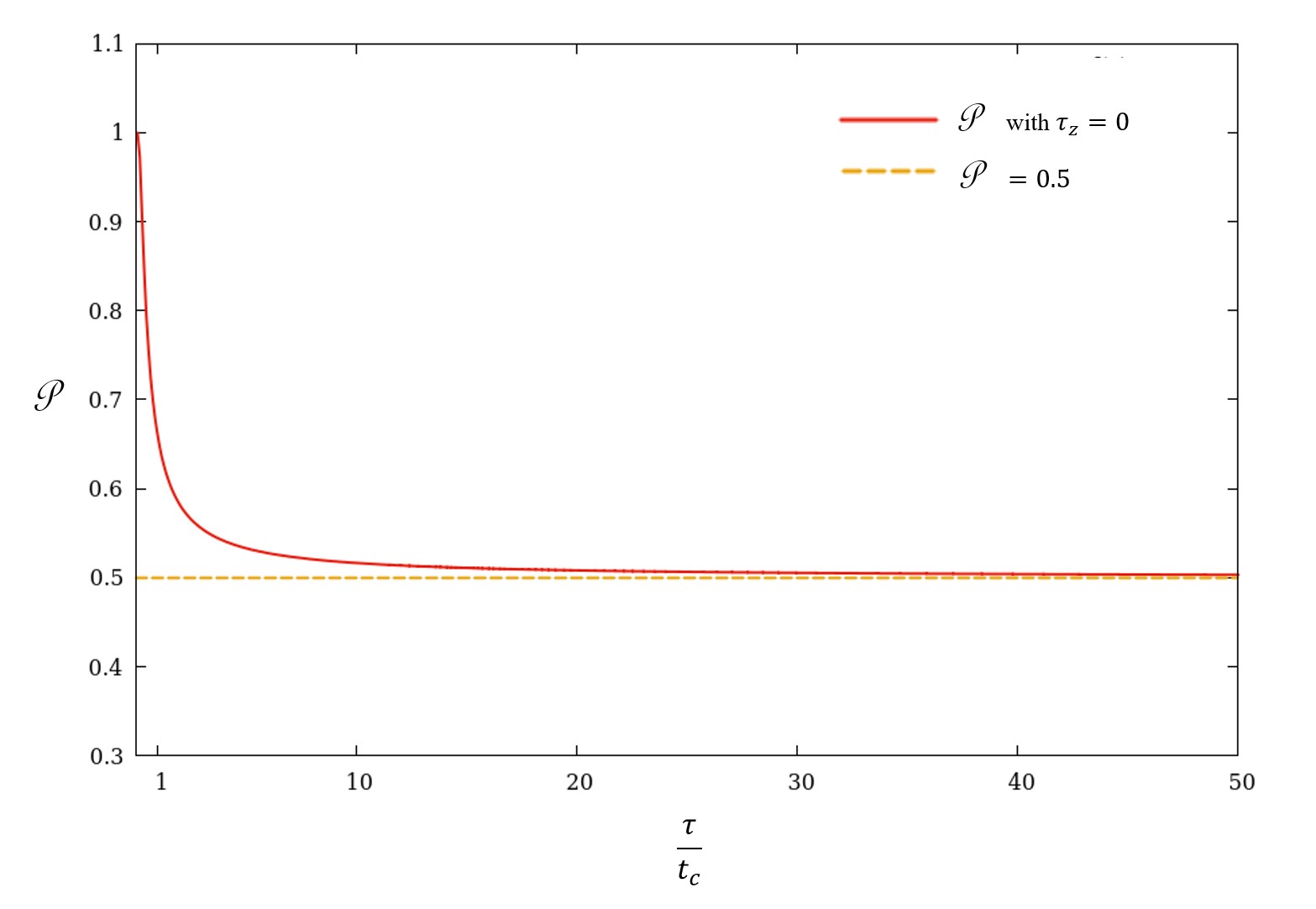}
	\caption{\footnotesize{ Propagation coefficient as a function of the ratio $\frac{\tau}{t_c}$ obtained using Eq. \ref{prop_eq}. The red curve refers to the case where the time constant $\tau_z=0$.}}
	\label{fig:PcoeffCStia}
\end{figure}

Taking the derivative of both sides of Eq. \ref{CFD_cs} we can find the expression for the propagation coefficient $\mathscr{P}$ in the CS-TIA case which is given by (see appendix \ref{app:A} for the detailed derivation):

\begin{equation}\label{propcsa}
\mathscr{P}=\frac{\partial T_{peak}}{\partial t_c}~,
\end{equation}

\begin{equation}\label{prop_eq}
\mathscr{P}=\frac{e^{\frac{t_c}{\tau}}(\tau-\tau_z)(\tau e^{\frac{t_c}{\tau}}-\tau-t_c)}{\tau (\tau+\tau_z)(e^{\frac{t_c}{\tau}}-1)^2}~.
\end{equation}
\\
The value of the propagation coefficient $\mathscr{P}$ as a function of the ratio $\frac{\tau}{t_c}$ is shown in Fig.~\ref{fig:PcoeffCStia}: for time constant $\tau$ of the electronics much greater of the average collection time $t_c$,  $\mathscr{P}$ approaches the value $\frac{1}{2}$. If we consider the presence of the zero with time constant $\tau_z$ in the transfer function \ref{eq:FdT}, we find that the propagation coefficient has a smaller value. For slow electronics $\tau_z$ can help to obtain a slightly better timing resolution. For very fast electronics the contribution of $\tau_z$ can usually be neglected since its zero goes to extremely high frequency.

The fact that the propagation coefficient for relatively slow electronics approaches $\frac{1}{2}$ is not surprising. Considering the discussion in \cite{riegler}, when the time constant is much greater than the duration of the signal, the electronics start behaving as a system able to measure accurately the \textit{time centroid} $t_{cog}$, defined in Eq. \ref{cog}. The intrinsic resolution $\sigma_{sens}$ is, in this case, given by the standard deviation of all the time centroids $\sigma_{t_{cog}}$.

\begin{equation}\label{cog}
t_{cog}=\frac{\int{I(t)\cdot t dt}}{\int I(t) dt}.
\end{equation}

The response of the CS-TIA is the pulse response $h(t)$ delayed by the centroid time
\begin{equation}
V_{out}(t)\approx h(t-t_{cog})~,
\end{equation}

and the peaking time of the voltage output becomes

\begin{equation}
T_{peak} \approx \tau + t_{cog}~.
\end{equation}

Considering rectangular pulses it means that the time centroid is exactly

\begin{equation}
t_{cog}=\frac{t_c}{2}~.
\end{equation}

The peaking time for a rectangular pulse of duration $t_c$ is then

\begin{equation}
T_{peak} \approx \tau + \frac{t_c}{2}~.
\end{equation}\\
\vspace{0.5cm}
Since the propagation coefficient $\mathscr{P}$ is given by \ref{propcsa}, the resolution of the system can be written as

\begin{equation}\label{criteria}
\sigma_{t_s}=\frac{\sigma_{t_c}}{2}~.
\end{equation}

In this special case, the CCT distribution and the distribution of all the time centroids are both rectangular distributions, with the difference that all the time centroids are exactly half the charge collection times $t_c$, which leads also to half the standard deviation. 
This suggests a possible criterion for which considering charge-sensitive electronics and 3D sensors with a population of current signals with different duration, the time resolution can be estimated by taking half the standard deviation of the charge collection time distribution. 
As an application of this theory to CS-TIA  we can show here a simulation performed by means of the TFBoost code~\cite{TFBoost_cit,tfboost_tcode}, where we make the convolution of the current signal from an ideal 3D sensor, as those in Fig.~\ref{fig:ideal3D1}, having a distance between the electrodes $d=20\mu m$ and saturation velocities equal to $v_e=0.1 \frac{\mu \text{m}}{\text{ps}}$ and $v_h=0.095 \frac{\mu \text{m}}{\text{ps}}$. The front-end electronics used in the simulation has the same characteristics of the Timespot1 ASIC \cite{Piccolo_2022}, corresponding to what listed in table~\ref{table1}.


\begin{table}[h] 
\centering
\begin{tabular}{|c|c|c|c|c|}
\hline
$R_f$  &$R_D$  & $C_{in}$ & $C_{f}$ & $C_{L}$ \\
\hline
$3M\Omega$ & $570k\Omega$ & $100fF$     & $5fF$      & $21fF$\\    
\hline
\end{tabular}
\hspace{0.7cm}
\begin{tabular}{|c|c|c|}
\hline
$g_m$  & $\tau$  & $\tau_z$  \\
\hline
$55\cdot 10^{-6} S$ & $10.8ns$ & $91ps$    \\    
\hline
\end{tabular}
	\caption{\footnotesize{CS-TIA values used in the simulation.}}
	\label{table1}
    \renewcommand\thetable{1}
\end{table}   

\begin{figure}[h]
	\centering
	\includegraphics[scale=0.25]{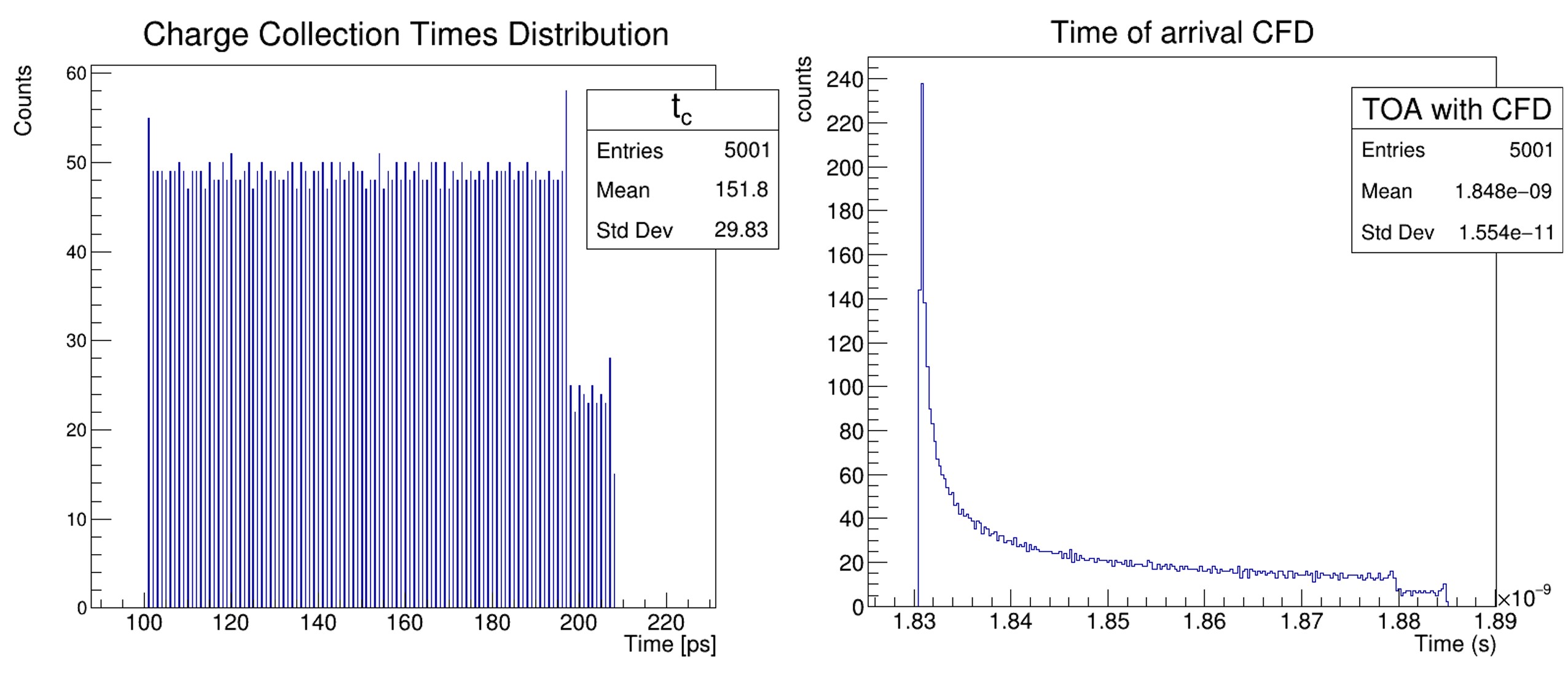}
	\caption{\footnotesize{CCT distribution of the currents used in the simulation (left), Time of arrival distribution obtained with TFBoost (right)}}
	\label{fig:tfboost_1}
\end{figure}

The results of the simulation are shown in Fig.~\ref{fig:tfboost_1}. We find a standard deviation of the CCT equal to $\sigma_{t_c}\sim 30ps$ and a resolution with a CFD with the threshold set at 35\% of the voltage peak equal to $\sigma_{t_s}\sim 15.5 ps$. Taking the ratio of the two standard deviations we find that the propagation coefficient is equal to
\begin{equation}
\mathscr{P}(\tau \sim 11ns) \sim 0.52.
\end{equation}

\subsubsection{Leading-edge time resolution in Fast-TIA}

\begin{figure}[h]
	\centering
	\includegraphics[scale=0.40]{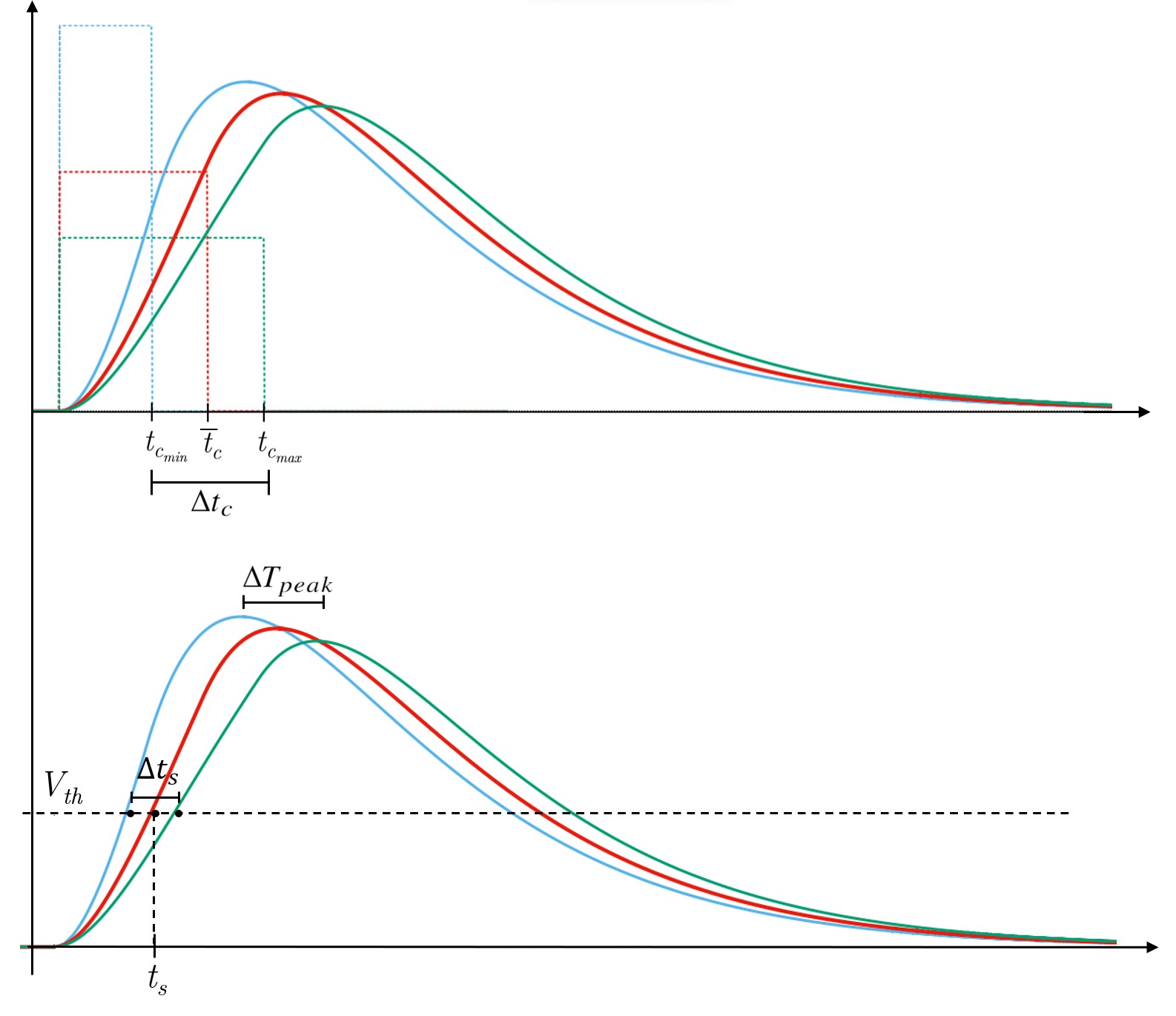}
	\caption{\footnotesize{Voltage signals with Fast-TIA for currents with different charge collection time $t_c$.}}
	\label{fig:plotTIA}
\end{figure}

The propagation coefficient calculated for the CS-TIA in the previous section increases if we have faster electronics with smaller time constant $\tau$ (Fig.~\ref{fig:PcoeffCStia}). This could induce thinking that a faster electronics would lead to a worse temporal resolution with respect to the CS-TIA case. Indeed this could be true, but only in specific cases. When the time constant starts being of the same order as the average duration of current pulse $t_c$ the resolution becomes dependent on the threshold position and the propagation coefficient $\mathscr{P}$ changes with it. The shape of the voltage signal changes for currents with different duration $t_c$ and we start suffering from ballistic deficit (Fig.~\ref{fig:plotTIA}). Signals with the same input charge will have variations of the voltage amplitude and, more important, the slope of the signal changes significantly. By decreasing the time constant $\tau$ the electronics is increasingly capable to follow the input currents and the peaking time $T_{peak}$ does not measure the time centroid anymore, but gets closer to measuring the duration of the current pulse $t_c$.

Considering the case when $ \tau \approx t_c$, we use now the expression of $ V_{out}(t)_{t<t_c} $ and set it equal to the chosen voltage threshold $V_{th}$. If we approximate the exponential to the second order we can solve for the threshold time $t_s$

\begin{equation}\label{eq:vrthle}
V_{th} = I_0R_{m_0}(1-e^{-\frac{t_{s}}{\tau}}(1+\frac{t_{s}}{\tau}))~,
\end{equation}
\begin{equation}\label{eq:vthapprox}
V_{th} \sim I_0R_{m_0}(1-(1-\frac{t_{s}}{\tau})(1+\frac{t_{s}}{\tau}))~,
\end{equation}
\begin{equation}\label{radice}
t_{s} \sim \tau \sqrt{\frac{V_{th}t_c}{Q_{in} R_{m_0}}}~.
\end{equation}

To calculate the propagation coefficient $\mathscr{P}$ we propagate the fluctuations finding
\begin{equation}\label{eq:sigmatFTIAlowthr}
\mathscr{P}(t_s)\sim \frac{1}{2}\frac{t_{s}}{t_c}~.
\end{equation}

The resolution at the threshold $V_{th}$ is

\begin{equation}\label{eq:sigmatFTIAlowthr}
\sigma_{t_s} =\mathscr{P}(t_s)\sigma_{t_c} ~.
\end{equation}

The time resolution $\sigma_{t_s}$ is now a function of the threshold time $t_s$. To understand what this means we can consider Fig.~\ref{fig:plotTIA}: our sensor is producing currents with rectangular shapes and different duration with a certain average $\overline{t_c}$; the current with average duration $\overline{t_c}$ will cross the threshold $V_{th}$ at time $t_s(\overline{t_c})$. Currents with shorter charge collection time will reach the threshold before time $t_s(\overline{t_c})$ while longer current pulse after time $t_s(\overline{t_c})$. If the charge collection times do not change in a large range we can assume that this variation is linear so that

\begin{equation}\label{eq:sigmatFTIAlowthr}
t_{s}(t_c) \sim t_s(\overline{t_c}) + \mathscr{P}(V_{th},\overline{t_c}) (t_c-\overline{t_c})~,
\end{equation}

the propagation coefficient can also be written using Eq. \ref{radice}
\begin{equation}\label{eq:sigmatFTIAlowthr}
\mathscr{P}(V_{th},\overline{t_c})\sim\frac{1}{ 2} \sqrt{\frac{V_{th}\overline{t_c}}{Q_{in} R_{m_0}}}~,
\end{equation}

where now we made explicit the dependence on the chosen voltage threshold. The time at the threshold $t_s$ and the propagation coefficient $\mathscr{P}$ grow as the square root of the voltage threshold $ V_{th} $. By setting $ V_{th} $ to a small value according to a $ t_s<t_c $ condition, we can reduce the sensor's intrinsic jitter. However, the approximation of equation~\ref{eq:vthapprox} tends to underestimate the fluctuations, being valid only for very small threshold values. For a more complete understanding, we can consider the time $ t $ as a function of $ t_c $ and derive both sides of the equation~\ref{eq:vrthle} as follows

\begin{equation}
 \frac{\partial}{\partial t_c}\bigg(\frac{V_{th}t_c}{Q_{in}R_{m_0}} \bigg)= \frac{\partial}{\partial t_c} \bigg(1-e^{-\frac{t_s}{\tau}}\big(1+\frac{t_s}{\tau}\big)\bigg)~,
\end{equation}
\begin{equation}
\frac{V_{th}}{Q_{in}R_{m_0}}=\frac{e^{\frac{t_s}{\tau}}t_s}{\tau^2} \frac{\partial t_s}{\partial t_c}~,
\end{equation}
solving for $ \frac{\partial t_s}{\partial t_c} $ we can write the derivative as\\
\begin{equation}
\frac{\partial t_s}{\partial t_c}=\frac{V_{th}}{V^{\prime}(t_s)t_c}~.
\end{equation}
\\
The resolution at the threshold becomes

\begin{equation}
\sigma_{t_s}=\frac{V_{th}}{V^{\prime}(t_s)t_c}\sigma_{t_c}~.
\end{equation}

Setting the threshold to the value corresponding to the maximum slope condition when $ t_s=\overline{t_c} $, we can use equations~\ref{eq:dVoutfast} and~\ref{eq:VoutfastmaxSl} and get the time resolution
\begin{equation}\label{key}
\sigma_{t_s}\approx{(e-2)}\sigma_{t_c}\sim 0.71\cdot\sigma_{t_c}~.
\end{equation}

The condition $ V_{th}=V_{\substack {\scriptscriptstyle Max\\Slope}} $ implies transporting about $ 70\% $ of the charge collection time fluctuations to the time resolution at the threshold. The total time resolution, in this case, could be dominated by the sensor contribution with respect to the electronic jitter that depends on the amount of charge generated (see section \ref{jitter}). Using equation~\ref{eq:vrthle} and the definition of the slope, the propagation coefficient becomes

\begin{equation}\label{eq:dtdtcle}
\mathscr{P}(t_s)=\frac{\tau^2(e^{\frac{t_s}{\tau}}-1)-\tau t_s}{t_s\cdot t_c}~.
\end{equation}

\begin{figure}[h]
	\centering
	\includegraphics[scale=0.37]{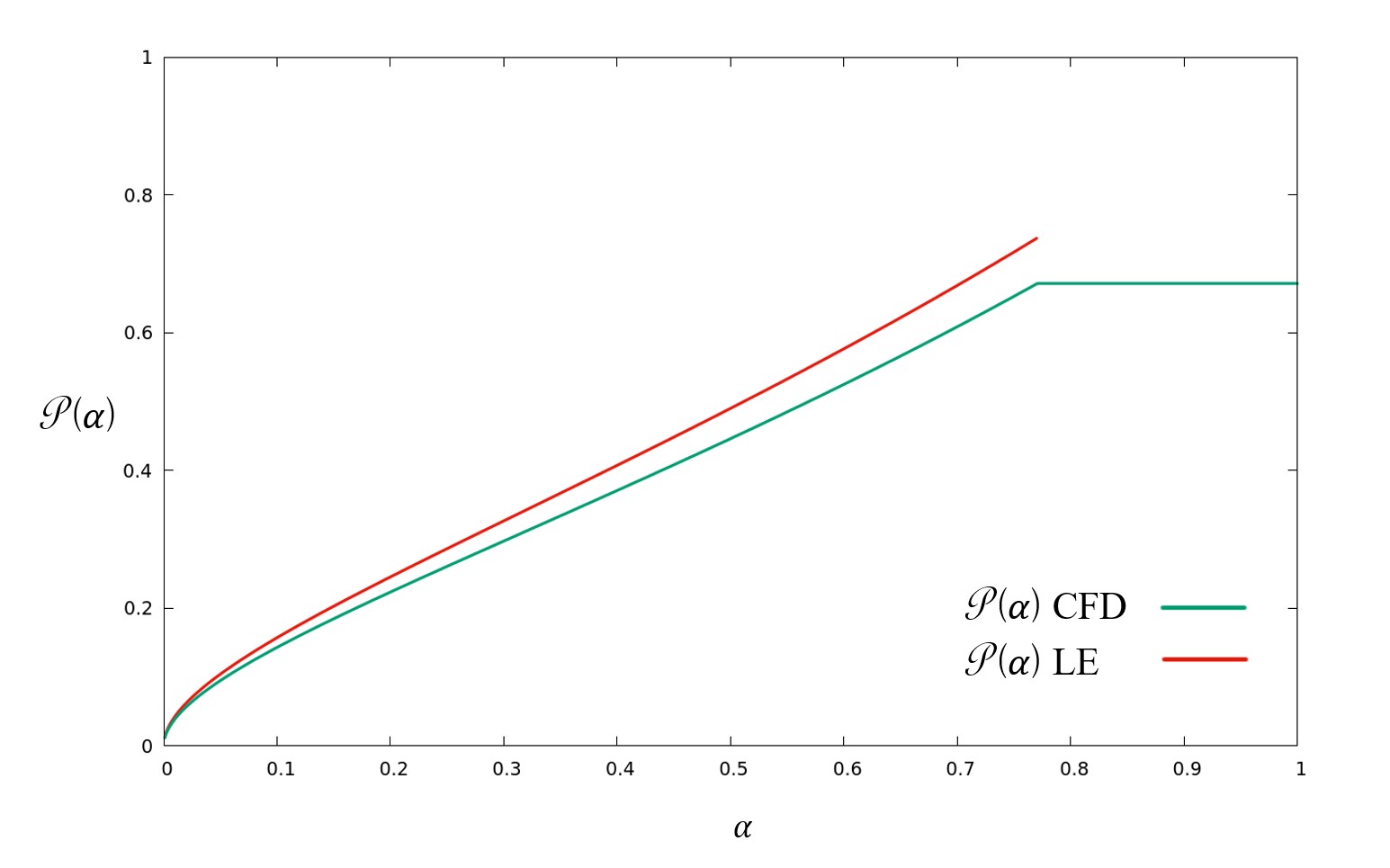}
	\caption{\footnotesize{Propagation coefficient $\mathscr{P}$ for the leading edge discrimination and for the constant fraction discrimination as a function of the threshold $\alpha$}}
	\label{fig:P_CFDvsLE}
\end{figure}

Fig.~\ref{fig:P_CFDvsLE} shows a plot of the propagation coefficient as a function of the fraction $\alpha$ of the signal for $t_c=\overline{t_c}$. For the LE, only the value of $\mathscr{P}$ for $t_s < \overline{t_c} $ are shown. A higher threshold could be also chosen but this would always lead to worse results in terms of intrinsic resolution. It must be also pointed out that if we choose a threshold higher than a certain value, with the leading edge discrimination we could lose some of the events with larger charge collection time (Fig.~\ref{fig:tfboost_1}). This happens when the chosen threshold is higher than the voltage reached at the maximum peaking time $T_{peak}(t_{c_{max}})$ relative to the maximum charge collection time $t_{c_{max}}$.

Even for fast electronics, the LE will be affected by time walk fluctuation, but they can be tolerable allowing us to still reach very good performances, as shown in \cite{JINST-TimeSpot}. Of course, best performances can be obtained if discrimination with time walk correction is used, such as the constant fraction method that will be treated in the next section.
 
\subsubsection{Constant-fraction time resolution in Fast-TIA}

To find the propagation coefficient $\mathscr{P}$ for the constant fraction case we consider the following equation
\begin{equation}\label{CFD_FT}
V_{out}(t_s)_{t<t_c}=\alpha V_{out}(T_{peak})~.
\end{equation}

The derivation is similar to the case for the CFD of the CS-TIA and is done in appendix \ref{app:B}. Similarly to the LE for the Fast-TIA we find that the propagation coefficient $\mathscr{P}$ is a function of the  time $ t_s $ that is the time at threshold fixed at the fraction $ \alpha $ of the voltage peak $ V_{peak}=V_{out}(T_{peak}) $. In particular,

\begin{equation}\label{cfdtiatc}
\mathscr{P}(t_s)_{t_s<t_c}=\frac{t_c}{\tau}\frac{e^{\frac{t_c}{\tau}}(\tau e^{\frac{t_s}{\tau}}-t_s-\tau)}{t_s(e^{\frac{t_c}{\tau}}-1)^2}~,
\end{equation}

while for $ t_s>t_c $, we find that the propagation coefficient $\mathscr{P}$  is independent of $t_s$ and is given by

\begin{equation}
\mathscr{P}_{t_s>t_c}=\frac{\partial T_{peak}}{\partial t_c}~,
\end{equation}

\begin{equation}\label{dtpeaktc}
\mathscr{P}_{t_s>t_c}=\frac{e^{\frac{t_c}{\tau}}(\tau e^{\frac{t_c}{\tau}}-\tau-t_c )}{\tau (e^{\frac{t_c}{\tau}}-1)^2}~.
\end{equation}\\

Eq. \ref{CFD_FT} can be solved numerically to find the time $ t_s $ as a function of the fraction $\alpha$ so that also the propagation coefficient can be expressed as a function of the fraction of the CFD. Fig.~\ref{fig:P_CFDvsLE} shows the propagation coefficient $\mathscr{P}(\alpha)$ that as expected is a growing function of the threshold. 

For fractions $\alpha$ higher than $V_{\substack {\scriptscriptstyle Max\\Slope}} $ the value of $\mathscr{P}$ stays constant and the front-end propagates about 70\% of the charge collection times fluctuation $\sigma_{t_c}$. Fig.~\ref{fig:P_CFDvsLE} shows also a comparison between LE and CFD discrimination for time constant $\tau \approx t_c$ where it can be seen that the two discrimination algorithms perform in a similar manner with fast electronics if we always deposit the same charge. If charge fluctuations are present, we will have an additional time walk of the time at threshold $t_s$ that can be eliminated with the constant fraction method or with a Leading Edge assisted with some kind of compensation (e.g. amplitude correction or TOT correction).

\subsubsection{Propagation coefficient \texorpdfstring{$\mathscr{P}$}{\textit{P}} for different \texorpdfstring{$\tau$}{tau}}
\begin{figure}[h]
\centering
    \hspace{1.5cm}
	\includegraphics[scale=0.25]{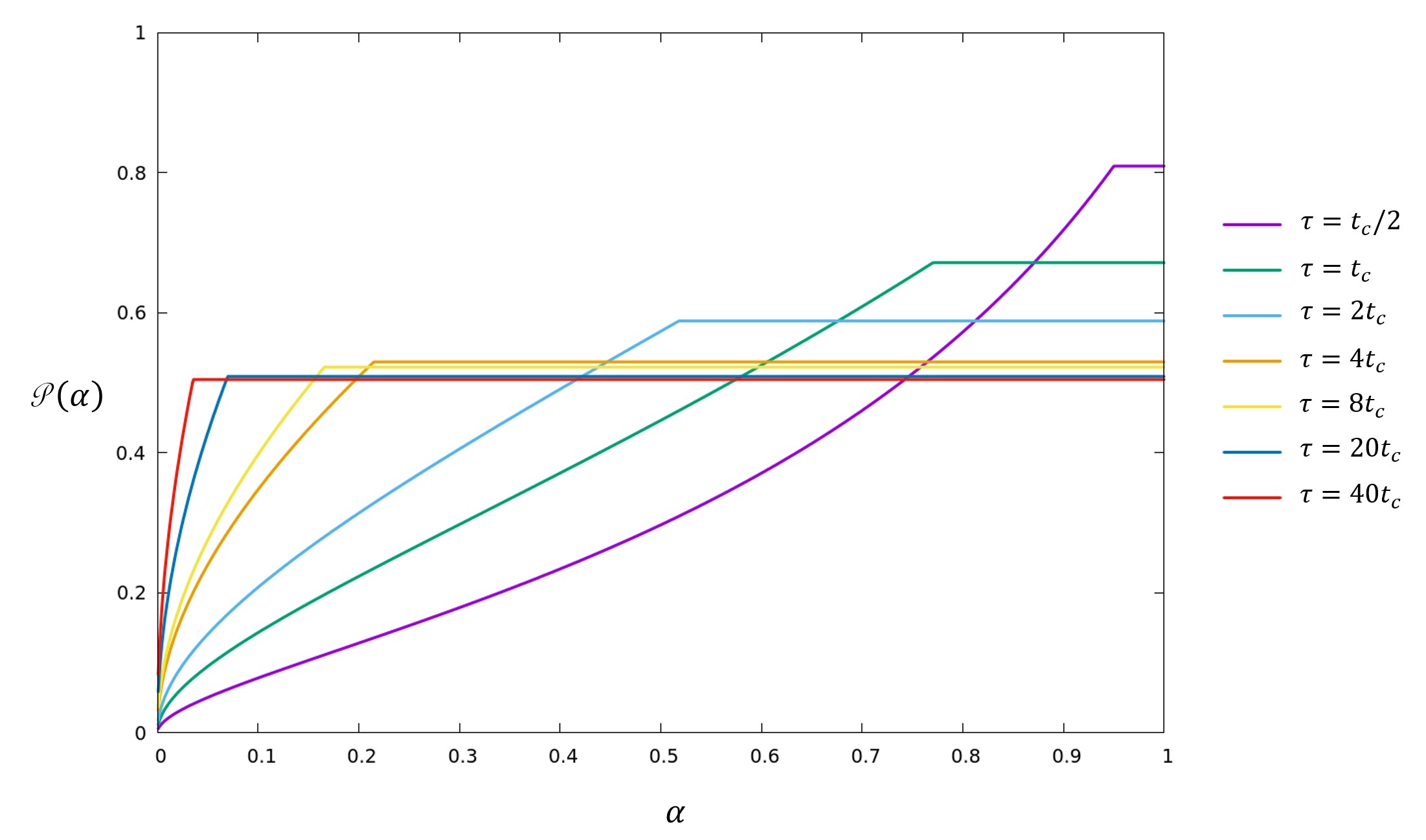}
	\includegraphics[scale=0.25]{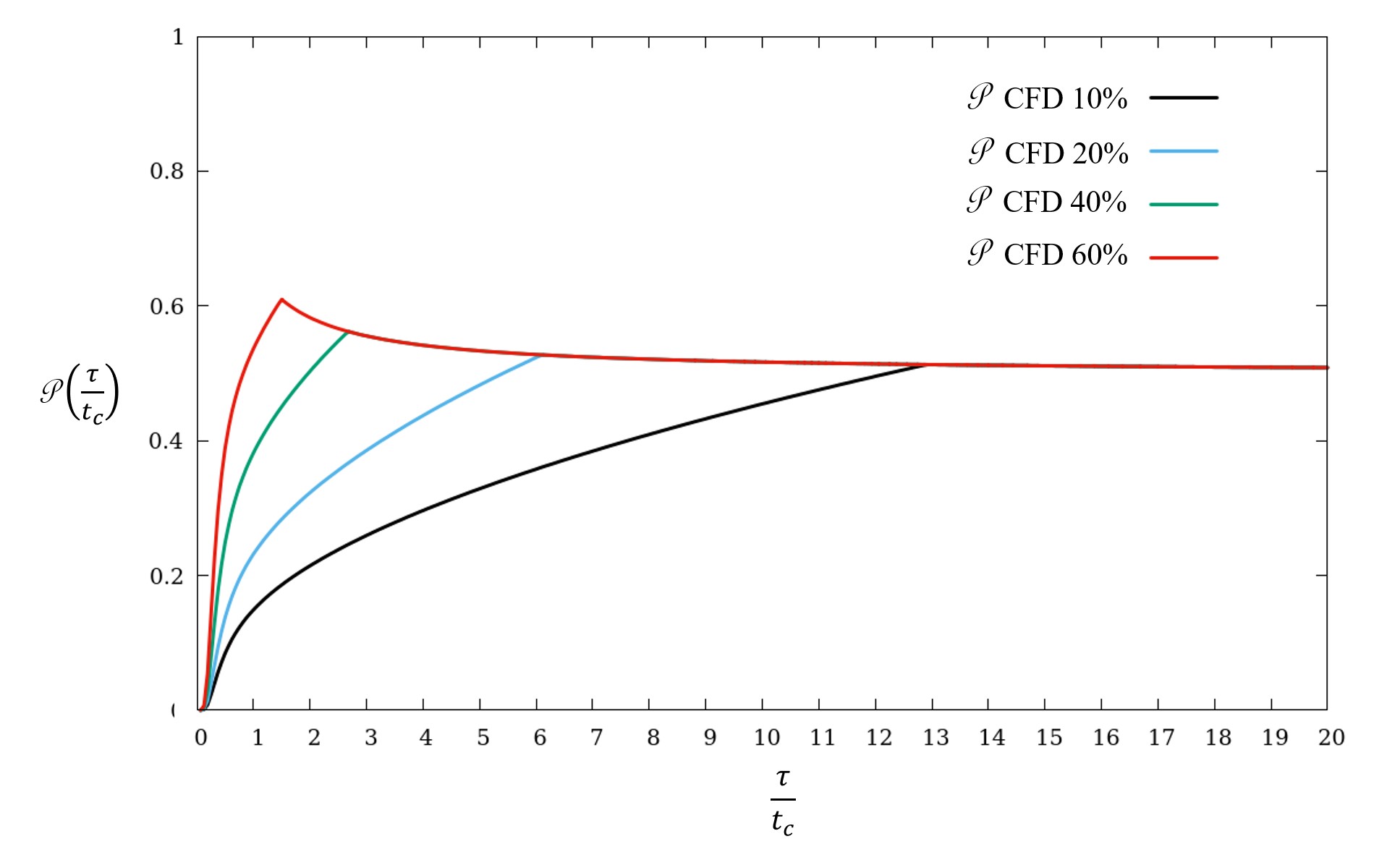}
	\caption{\footnotesize{Propagation coefficient $\mathscr{P}$ for the constant fraction discrimination as a function of the threshold $\alpha$ for different time constant $\tau$ (top), Propagation coefficient $\mathscr{P}$ for the constant fraction discrimination as a function of the ratio $\frac{\tau}{t_c}$ for different threshold $\alpha$ (bottom)}}
	\label{fig:Pvsalpha}
\end{figure}

Up to now we have considered the value that the propagation coefficient $\mathscr{P}$ assumes in two particular cases, namely the CS-TIA where the time constant $ \tau \gg  t_c $ and the Fast-TIA in which $ \tau \approx t_c $. However, as Eq. \ref{omegan} and \ref{tau_FT} shows, the value of the time constant $ \tau $ depends on several factors: the capacitances involved, through the quantity $ \xi $, in particular the capacitance $ C_D $ of the sensor that can be the dominant one in certain configurations; the value of the DC trans-impedance and finally the trans-conductance $ g_m $ of the transistor. The trans-conductance $g_m$ is often the most limiting factor since is directly related to the power consumption and, in circuits with very strict power constraints it could be impossible to reach the value of the time constant $\tau$ of the same order of the average duration of the current pulse. To understand what happens if more speed of the electronics, could be exploited, Fig.\ref{fig:Pvsalpha} (top) the value of the propagation coefficient $\mathscr{P}$ is shown as a function of the threshold $\alpha$ for different time constant $\tau$. To take advantage of the reduction of the propagation coefficient as a function of the threshold, the time constant can be several times greater than the average charge collection time $t_c$ provided that a high SNR is given.  As an example, let us consider a system with an SNR$=50$; setting the threshold to $5\sigma_v$ to avoid false hits, the threshold would be at 10\% of the voltage peak. If we look at Fig.\ref{fig:Pvsalpha} (bottom), that shows the propagation coefficient $\mathscr{P}$ as a function of the ratio $\frac{\tau}{t_c}$, with a time constant five times greater than the average collection time $t_c$, we still be able to halve the propagation coefficient $\mathscr{P}$ reaching an intrinsic time resolution $\sigma_{t_s}=\sigma_{t_c}/4$ instead of half $\sigma_{t_c}$.

\section{Contributions to time resolution for a real 3D trench detector}

This section summarizes the two contributions to the time resolution of real 3D trench sensors, as seen in Eq. \ref{eq:sigmast2}. The intrinsic contribution of a real sensor can be quite different from what is seen in the ideal case and requires accurate simulations to obtain the induced currents. The contribution due to the electronic jitter is instead reported using the known equations.

\subsection{Intrinsic time resolution of real 3D trench sensor}

\begin{figure}[h]
\centering
    \includegraphics[scale=0.6]{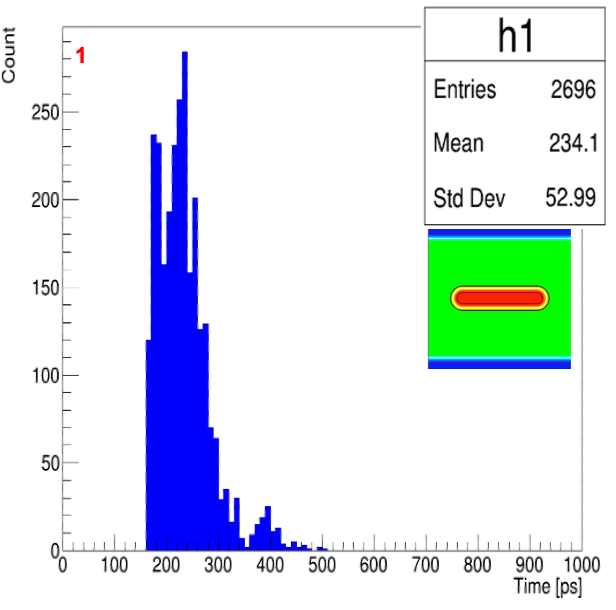}
	\caption{\footnotesize{CCT distribution obtained with a TCoDe simulation of the TimeSPOT sensor \cite{TCode,tcode2}. }}
	\label{fig:CCT_TCODE}
\end{figure}

A description of a 3D sensor using an ideal 3D parallel plate geometry helps to understand one of the most important contributions to the intrinsic resolution of this type of sensors, namely, the fact that the currents generated have different durations due to the different drift times of the carriers depending of the position of the impinging particle. This leads to obtaining a distribution of these durations $t_c$ which is the CCT distribution. In a real 3D detector there are other aspects that can affect the duration and the shape of the signals, in particular, diffusion and variation of the electric and weighting fields. When all the effects are considered, the shape and duration of the currents inside the detector can be quite different to what we would expect considering an ideal 3D parallel plate geometry with carriers at saturation velocities and in a constant weighting field. Indeed, a great effort has been made in order to optimize the geometry of this type of sensors in order to optimize their performance \cite{TCode,tcode2}. Fig. \ref{fig:CCT_TCODE} shows the CCT distribution obtained with a simulation with the software TCoDe \cite{TCode,tcode2} of the 3D detector developed within the TimeSPOT project \cite{timespot}.

The distribution shows a standard deviation $\sigma_{t_c}\sim 53$ ps. Considering the detector geometry, the distance between electrodes and the saturation velocities used previously in section \ref{subsec:SensorJitter}, we would expect a value of $\sigma_{t_c}^{Ideal}\sim 36$ ps. Therefore, also in an strongly optimized detector the value of $\sigma_{t_c}$ is indeed dominated by the fact that we have different duration because of the different drift distances that the carriers have to cover but all the other contributions (i.e. weighting field, diffusion) could worsen severely the spread of the duration of the currents. The propagation coefficient $\mathscr{P}$ defined in the previous section is still useful also for CCT distributions such as Fig.\ref{fig:CCT_TCODE} and the criterion found for the CS-TIA case (Eq. \ref{criteria}) still holds. In particular, supposing to process the currents of the detector with the CCT shown in Fig.\ref{fig:CCT_TCODE}, we would expect to obtain an intrinsic resolution of about
\begin{equation}
\sigma_{t_s}^{\footnotesize CS-TIA }\sim \frac{\sigma_{t_c}}{2} = 26\,\text{ps}.
\end{equation}
This has also been confirmed by extensive simulations made with the software packages TCoDe and TFBoost \cite{TCode,tcode2,tfboost_tcode}. If a Fast-TIA is used instead with a time constant $\tau \approx t_c$, the resolution could be strongly improved depending of the noise present in the system. If a very low threshold can be used, the intrinsic contribution could be heavily reduced as seen both in experiment \cite{JINST-TimeSpot}, and simulations \cite{Brundu_2021}, where an intrinsic resolution of about $15ps$ was estimated for the TimeSPOT sensor. Considering the value of $\sigma_{t_c}$ found in the TCoDe simulation, using fast electronics and a low threshold allow to obtain a propagation coefficient $\mathscr{P} \sim 0.28$ that is consistent with the threshold fraction used in \cite{JINST-TimeSpot}.
 
\subsection{Front-end electronics jitter}\label{jitter}

Conceptually, the contribution of the front-end electronics to the system time resolution can be interpreted as the projection of the electronic noise onto the time axis and is defined as electronic time jitter or $\sigma_{\rm ej}$, already found in Eq.~\ref{eq:sigmast}. This is given by~\cite{Spieler}, 

\begin{equation}
\sigma_{\rm ej}=\sigma_v \Bigg(\frac{dV}{dt}\Bigg)^{-1}~.
\label{eq:sigmaj}
\end{equation} 

To evaluate $\sigma_{\rm ej}$ we need both the time derivative of the signal function $V_{out}(t)$ and the voltage noise $\sigma_v$ \footnote{Explicit calculation of the noise performance of the two configurations can be found in \cite{FastTiming}.}. By looking at the expressions for the slope of the voltage signal in the two cases (Eq. \ref{eq:dVdtCSA} and Eq. \ref{eq:dVoutfast}), we see that is directly proportional to the total charge and trans-conductance $g_m$ and inversely proportional to the quantity $\xi$ (Eq. \ref{xidef}). The obtainable jitter depends strictly by the amount of charge available, the power constraints (i.e. the $g_m$, from which also depends the noise performance) and all the capacitance involved in the circuit the are crucial to the final speed of the electronics.

As shown in \cite{blum2008particle}, the electronic jitter performance improves if the time constant is of the same order of the duration of the current pulse. For planar detectors, this means a trade-off between intrinsic contribution to time resolution, (that gets smaller for electrodes closer to each other), and electronic jitter which is inversely proportional to the total charge (that for planar detector is proportional to the sensor thickness).

The 3D detectors, instead, allow to decouple electrodes distance and produced charge so that an higher signal to noise ratio can be obtained and very good performance in terms of time resolution can be reached, even without benefiting of a gain charge mechanism \cite{JINST-TimeSpot,CardiniPisa,Andreaiworid}. 

Considering the two electronics analyzed, we can express the electronic jitter with the following relations (more details can be found in appendix \ref{app:C})
 
 \begin{equation}
\sigma_{\rm ej}^{CS}\sim \frac{\sigma_v \xi}{\overline{Q_{in}}g_m}~.
\label{sig_cs}
\end{equation} 

Here the term $\overline{Q_{in}}$ refers to the average amount of charge deposited. 

Another way to express the jitter is in term of the signal to noise ratio SNR and time constant of the system $\tau$,
 \begin{equation}
\sigma_{\rm ej}^{CS}\sim \frac{1}{SNR}\frac{\tau}{e}~. 
\end{equation} 
For the Fast-TIA case, the jitter at the maximum derivative is
 \begin{equation}
\sigma_{\rm ej}^{F}\sim \frac{0.96}{SNR}\cdot\tau~.
\end{equation} 
Using a Fast-TIA that has time constant $\tau\sim 200ps$ of the same order of the average duration of the currents in the 3D detector, we find that with SNR$=20$ the electronic jitter $\sigma_{ej}$ would be less than 10 ps.

\section{Conclusions}
\label{sec:discussion}

Starting from the Eq. \ref{eq:sigmast2}, that is completely general for many types of sensors, we have found an explicit description of the two contributions $\sigma_ {\rm sens}$ and $\sigma_{ej}$ in the case we have a detector that produce currents with different durations, such as the 3D silicon sensor with parallel plate geometry (\textit{trench} structure) connected to a feedback TIA electronics. 
The final performance in terms of time resolution depends strongly on the characteristic of the system detector+electronics, that can be characterized by a time constant $\tau$, that depends on the electronics itself (i.e. trans-conductance $g_m$, feedback network, capacitances etc. ) and sensor capacitance $C_D$.\\
\\
The time resolution achievable with such system can be summarized in two cases:

\subsection*{CS-TIA ($\tau \gg  t_c$)}
The time constant of the system $\tau$ is much greater than the average current duration $t_c$. The intrinsic contribution of the detector is independent of the discrimination threshold value and is given by
\begin{equation}
    \sigma_ {\rm sens}= \mathscr{P} \sigma_{t_c}~,
\end{equation}

where $\mathscr{P}$ is the \textit{timing propagation coefficient} and $\sigma_{t_c}$ is the standard deviation of the CCT distribution. For 3D trench detector we have that

\begin{equation}
    \mathscr{P} \approx 0.5~,
\end{equation}

\begin{equation}
    \sigma_{t_c} \sim \frac{t_c^{max}-t_c^{min}}{\sqrt{12}}~,
\end{equation}
where $t_c^{max}$ and $t_c^{min}$ are the maximum and minimum charge collection time of the currents in the detector.
With short inter-electrode distance, $t_c^{max}$ and $t_c^{min}$, and, consequently, $\sigma_{t_c}$ can be made very small, still having enough charge since 3D detectors have thickness and distance of electrodes decoupled.\\

If we introduce also the electronic jitter for the charge-sensitive case, $\sigma_{ej}^{CS}$, the final time resolution can be expressed as

\begin{equation}
    \sigma_{t}= \sqrt{ \Big(\mathscr{P} \sigma_{t_c}\Big)^2 + \Big(\sigma_{ej}^{CS}\Big)^2}~.
\end{equation}

If we can set the threshold very low becomes
\begin{equation}
    \sigma_{t} \sim \sqrt{ \Big(\frac{\sigma_{t_c}}{2} \Big)^2 + \Big(\frac{1}{SNR}\frac{\tau}{e}\Big)^2}~,
\end{equation}

where $\tau$ is the time constant of the system and SNR is the signal to noise ratio.
\subsection*{Fast-TIA ($\tau \sim t_c$)}

The time constant of the system is of the same order of the average duration of the currents. The intrinsic contribution of the detector is now dependent on the chosen threshold. This can still be expressed using the propagation coefficient that is a growing function of the threshold $\alpha$. Referring to the constant fraction method the intrinsic contribution is

\begin{equation}
    \sigma_ {\rm sens}(\alpha)= \mathscr{P}(\alpha) \sigma_{t_c}~.
\end{equation}

Also the electronic jitter for the Fast-TIA $\sigma_{ej}^{F}$ is threshold dependent, being minimum at about 75\% of $V_{peak}$ and growing from lower threshold. The standard deviation $\sigma_ {\rm sens}$ gets smaller for lower threshold but the jitter increases, which means that exist an optimum value of the threshold that minimize the time resolution. The total time resolution can be written as

\begin{equation}
    \sigma_{t}(\alpha)= \sqrt{ \Big(\mathscr{P}(\alpha) \sigma_{t_c}\Big)^2 + \Big(\sigma_{ej}^{F}(\alpha)\Big)^2}~.
\end{equation}
\\
A reasonable estimate with threshold at about $25\%$ of $V_{peak}$ is
\begin{equation}
    \sigma_{t}(\alpha=25\%) \sim \sqrt{ \Big( \frac{\sigma_{t_c}}{4} \Big)^2 + \Big(\frac{1.2~\tau}{SNR}\Big)^2}~.
\end{equation}

\acknowledgments

This work was supported by the Fifth Scientific
Commission (CSN5) of the Italian National
Institute for Nuclear Physics (INFN), within the Project TimeSPOT. 
The authors wish to thank Angelo Loi for his help in providing the plots used in Fig.~\ref{fig:CCT_TCODE} and Davide Brundu for the useful discussions.

\appendix
\section{Propagation coefficient $\mathscr P$ for the Constant Fraction Discrimination\\ case:$~~t_s>t_c$}
\label{app:A}

Let us consider first the CS-TIA case with the voltage signal for $t>t_c$. The threshold time $t_s$ is the one that satisfied the following equation

\begin{equation}\label{eq_app1}
 e^{-\frac{t_{s}}{\tau}}\Bigg(B\frac{t_{s}}{\tau} + C \Bigg)=\alpha \Bigg[  e^{-\frac{T_{peak}}{\tau}}\Bigg(B\frac{T_{peak}}{\tau} + C \Bigg) \Bigg]~,
\end{equation}

where $\alpha$ indicate the fraction of the voltage peak.
\vspace{0.5cm}\\
Taking the derivative of both sides with respect to $t_c$ we obtain
\begin{equation}
V^{'}(t_s)\frac{\partial t_s}{\partial t_c} + e^{-\frac{t_{s}}{\tau}} \Bigg( \frac{\partial B}{\partial t_c}\frac{t_s}{\tau} + \frac{\partial C}{\partial t_c} \Bigg) = \alpha \Bigg[  e^{-\frac{T_{peak}}{\tau}} \Bigg( \frac{\partial B}{\partial t_c}\frac{T_{peak}}{\tau} + \frac{\partial C}{\partial t_c} \Bigg)                         \Bigg]~,
\end{equation}

where $V^{'}(t_s)$ is the normalized derivative (i.e. obtained dividing for $I_0R_{m_0}$).\\
\\
Solving for $\frac{\partial t_s}{\partial t_c}$,
\begin{equation}
\frac{\partial t_s}{\partial t_c}  =  \frac{\alpha\Bigg[  e^{-\frac{T_{peak}}{\tau}}\Bigg( \frac{\partial B}{\partial t_c}\frac{T_{peak}}{\tau}  + \frac{\partial C}{\partial t_c} \Bigg) \Bigg]-  e^{-\frac{t_s}{\tau}}\Bigg( \frac{\partial B}{\partial t_c}\frac{t_s}{\tau}  + \frac{\partial C}{\partial t_c}  \Bigg)}{V^{'}(t_s)}~,
\end{equation}\\
and the derivative $V^{'}(t_s)$ can be written,
\begin{equation}
V^{'}(t_s)=  e^{-\frac{t_{s}}{\tau}} \frac{B}{\tau}\Bigg( \frac{T_{peak}-t_s}{\tau} \Bigg)~,
\end{equation}

substituting $\alpha$ using Eq. \ref{eq_app1} we find:
\begin{equation}
\frac{\partial t_s}{\partial t_c}  =  \frac{\Bigg(B\frac{t_{s}}{\tau} + C \Bigg)\Bigg( \frac{\partial B}{\partial t_c}\frac{T_{peak}}{\tau}  + \frac{\partial C}{\partial t_c} \Bigg)-\Bigg(B\frac{T_{peak}}{\tau} + C \Bigg)\Bigg( \frac{\partial B}{\partial t_c}\frac{t_s}{\tau} + \frac{\partial C}{\partial t_c} \Bigg) }{\Bigg(B\frac{T_{peak}}{\tau} + C \Bigg) \frac{B}{\tau}\Bigg( \frac{T_{peak}-t_s}{\tau} \Bigg)}
\end{equation}

some of the terms cancel out and the propagation coefficient becomes independent from the threshold time $t_s$
\begin{equation}
\frac{\partial t_s}{\partial t_c}  = \tau \frac{\Bigg( C \frac{\partial B}{\partial t_c}-B \frac{\partial C}{\partial t_c} \Bigg)}{B^2}~,
\end{equation}

since $T_{peak}$ is given by

\begin{equation}
T_{peak}=\tau \frac{B-C}{B}~,
\end{equation}

taking the derivative of $T_{peak}$ with respect to $t_c$ we find that
\begin{equation}
\frac{\partial t_s}{\partial t_c}=\frac{\partial T_{peak}}{\partial t_c}~.
\end{equation}

We conclude that the propagation coefficient for $t>t_c$ is given by the derivative of the peaking time $T_{peak}$ with respect to the charge collection time $t_c$
\begin{equation}
\mathscr{P}=\frac{\partial T_{peak}}{\partial t_c}
\end{equation}

\section{Propagation coefficient $\mathscr P$ for the Constant Fraction Discrimination\\ case:$~~t_s<t_c$}
\label{app:B}

Let's consider first the Fast-TIA case with the voltage signal for $t<t_c$. The threshold time $t_s$ is the one that satisfied the following equation

\begin{equation}\label{eq_app2}
\Bigg(1-e^{-\frac{t_{s}}{\tau}} \Bigg( 1+ \frac{t_s}{\tau}\Bigg) \Bigg)=\alpha \Bigg[  e^{-\frac{T_{peak}}{\tau}}\Bigg(B\frac{T_{peak}}{\tau} + C \Bigg) \Bigg]~.
\end{equation}\\

Taking the derivative of both sides with respect to $t_c$ we obtain
\begin{equation}
V^{'}(t_s)\frac{\partial t_s}{\partial t_c} = \alpha \Bigg[  e^{-\frac{T_{peak}}{\tau}} \Bigg( \frac{\partial B}{\partial t_c}\frac{T_{peak}}{\tau} + \frac{\partial C}{\partial t_c} \Bigg)                         \Bigg]~,
\end{equation}

again $V^{'}(t_s)$ is the normalized derivative (i.e. obtained dividing for $I_0R_{m_0}$).\\
\\
Solving for $\frac{\partial t_s}{\partial t_c}$,

\begin{equation}
\frac{\partial t_s}{\partial t_c}  = \alpha \frac{\Bigg[  e^{-\frac{T_{peak}}{\tau}}\Bigg( \frac{\partial B}{\partial t_c}\frac{T_{peak}}{\tau}  + \frac{\partial C}{\partial t_c} \Bigg) \Bigg]}{V^{'}(t_s)}~,
\end{equation}\\

and substituting $\alpha$ using Eq. \ref{eq_app2} we find
\begin{equation}
\frac{\partial t_s}{\partial t_c}  =  \frac{\Bigg(1-e^{-\frac{t_{s}}{\tau}} \Bigg( 1+ \frac{t_s}{\tau}\Bigg) \Bigg)\Bigg( \frac{\partial B}{\partial t_c}\frac{T_{peak}}{\tau}  + \frac{\partial C}{\partial t_c} \Bigg)}{ \Bigg(e^{-\frac{t_{s}}{\tau}}\frac{t_s}{\tau^2} \Bigg)\Bigg(B\frac{T_{peak}}{\tau} + C \Bigg)}~,
\end{equation}

using the value of $T_{peak}$ for the Fast-TIA,

\begin{equation}
T_{peak}=\frac{e^{\frac{t_c}{\tau}}t_c}{e^{\frac{t_c}{\tau}}-1}~,
\end{equation}\\

we find that for $t<t_c$ the derivative $\frac{\partial t_s}{\partial t_c}$ is equal to

\begin{equation}
\frac{\partial t_s}{\partial t_c}  =  \frac{t_c}{\tau} \frac{e^{\frac{t_c}{\tau}}(\tau e^{\frac{t_s}{\tau}}-t_s-\tau)}{t_s(e^{\frac{t_c}{\tau}}-1)^2}~.
\end{equation}

The propagation coefficient is in this case dependent of the chosen threshold $t_s$, and gets smaller for lower fraction $\alpha$:

\begin{equation}
\mathscr{P}(t_s)  =  \frac{t_c}{\tau} \frac{e^{\frac{t_c}{\tau}}(\tau e^{\frac{t_s}{\tau}}-t_s-\tau)}{t_s(e^{\frac{t_c}{\tau}}-1)^2}
\end{equation}

\section{Jitter approximation for CS-TIA and Fast-TIA}
\label{app:C}

\subsection{CS-TIA}
  \begin{equation}
\sigma_{\rm ej}^{CS}\sim \frac{\sigma_v \xi}{\overline{Q_{in}} g_m} 
\end{equation} 

using the expression of $\tau$ in Eq.\ref{omegan} we have

\begin{equation}
\sigma_{\rm ej}^{CS}\sim \frac{\sigma_v \tau^2}{\overline{Q_{in}} R_f} ~;
\end{equation} 

using the value of $V_{peak}$ in Eq. \ref{VpeakCS},

\begin{equation}
\sigma_{\rm ej}^{CS}\sim \frac{\sigma_v \tau}{\overline{Q_{in}} R_f} \tau = \frac{\sigma_v e \tau}{\overline{Q_{in}} R_f} \frac{\tau}{e}= \frac{\sigma_v }{V_{peak}}\frac{\tau}{e}=\frac{1}{SNR}\frac{\tau}{e}~,
\end{equation} 

\begin{equation}
\sigma_{\rm ej}^{CS}\sim\frac{1}{SNR}\frac{\tau}{e}
\end{equation} 

\subsection{Fast-TIA}

First we calculate the jitter with the threshold set to the maximum slope condition. Using Eq. \ref{eq:dVoutfast} we have

\begin{equation}
\sigma_{\rm ej}^{F}\sim \frac{\sigma_v e \tau^2}{\overline{Q_{in}} R_{m_0}}~;
\end{equation} 

considering the $V_{peak}$ expression we can write,

\begin{equation}
\sigma_{\rm ej}^{F}\sim \frac{\sigma_v e \tau^2}{\overline{Q_{in}} R_{m_0} } \cdot \frac{t_c \cdot e^{-\frac{T_{peak}}{\tau}}B}{t_c \cdot e^{-\frac{T_{peak}}{\tau}}B}~,
\end{equation} 

with $B=e^{\frac{t_c}{\tau}}-1$, $\frac{\overline{Q_{in}}}{t_c}=I_0$, $T_{peak}$ given by Eq. \ref{Tpeak_fast} and being $\tau=t_c$ we find

\begin{equation}
\sigma_{\rm ej}^{F}\sim \frac{\sigma_v }{V_{peak}} \Big(e(e-1)e^{-\frac{e}{e-1}} \Big)\tau~,
\end{equation} 

\begin{equation}
\sigma_{\rm ej}^{F}\sim \frac{0.96 }{SNR} \tau
\end{equation} 

\bibliographystyle{ieeetr}  
\bibliography{bibl.bbl}  

\end{document}